\documentclass[11pt,letterpaper]{article}
\usepackage[margin=0.5in]{geometry}
\usepackage{amssymb}
\usepackage{amsmath}
\usepackage{pifont}
\usepackage{array}
\usepackage{wrapfig}
\usepackage{graphicx}
\usepackage{url}
\usepackage{caption}
\usepackage{subcaption}
\usepackage{tablefootnote}
\usepackage[table,xcdraw]{xcolor}
\usepackage{cite}

\usepackage[labelfont=bf,figurename=Figure]{caption}

\title{Design of Memristive Lightweight Encryption For In-Memory Image Steganography \thanks{Preprint Submitted to arXiv}}

\author{Seyed Erfan Fatemieh$^{1,}$\thanks{Corresponding author: Seyed Erfan Fatemieh (erfanfatemieh@eng.ui.ac.ir)}~, Reza Shahdi Alizadeh$^{2,}$\thanks{Reza Shahdi Alizadeh (r.sh.alizadeh@gmail.com)}~, and Esmail Zarezadeh $^{3,}$\thanks{Esmail Zarezadeh (zarezadeh@aut.ac.ir)}}

\begin{document}
\maketitle

{\noindent$^{1}$ \small \textit{Department of Computer Architecture, Faculty of Computer Engineering, University of Isfahan, Isfahan 8174673441, Iran}}
{\noindent$^{2}$ \small \textit{Department of Computer Engineering, Yadegar-e-Imam Khomeini (RAH) Shahre Rey Branch, Islamic Azad University, Tehran, Iran\\}}
{\noindent$^{3}$ \small \textit{Department of Electrical Engineering, AmirKabir University of Technology, Tehran, Iran}}

\section*{Abstract}
With the expansion of data-intensive applications and increasing data volumes, providing an efficient solution to address growing energy consumption and performance degradation caused by the transfer of large amounts of data between the processor and the main memory has become a severe challenge. The frequent transfer of large amounts of data between internal chip units, memories, and their interconnections exacerbates the vulnerability of the data being accessed. Employing a memristive Computation In-Memory-Array (CIM-A) architecture limits data transfer, thereby addressing both challenges. Furthermore, by integrating lightweight cryptography, developed to secure data in hardware-constrained devices, with CIM-A architectures, the security of data in transit, especially across interconnections, can be ensured. This paper implements two standard lightweight stream ciphers, Trivium and Grain-128a, for CIM using stateful material implication (IMPLY) logic to address these combined security and performance challenges. In addition to redesigning the cryptographic structures, we reduce the hardware complexity of conventional IMPLY-based implementations by proposing an efficient method for shifting data within the shift registers. Applying the proposed data-shifting method to the registers of these ciphers reduces the number of computational steps and decreases energy consumption by up to 42\% and 44\%, respectively, compared to conventional implementations. Finally, the performance of the proposed circuits is evaluated in a steganography application, demonstrating their practical efficiency.

\subsection*{Keywords}
Computation In-Memory (CIM), Memristor, Lightweight Cryptography, IMPLY Logic, Steganography, Emerging Technologies.

\section{Introduction} \label{sec1}
Today, data is one of the most important components of computation, and its fast, energy-efficient processing has become a challenge for computer architects more than ever. The Internet of Things (IoT) and Artificial Intelligence (AI) are among the most important processing applications, expanding in various dimensions every day. From the perspective of an architect of a high-volume data-processing system, the secure transmission of data, alongside fast, and energy-efficient processing, is of particular importance. Maintaining data security and integrity poses numerous challenges, and using appropriate cryptographic mechanisms for secure data transmission, especially on edge devices and in IoT-related processing applications, is of high importance. The constant data movement between the processor and memory in the conventional von Neumann architecture is a major concern for designers and researchers, both in terms of resource consumption and circuit evaluation criteria, and in terms of the security of data accessed during inter-unit transfers, especially in interconnections \cite{ref1, ref2}.

By the middle of the first decade of the new century, technological improvements and reductions in chip transistor dimensions had significantly improved performance and addressed various processing needs \cite{ref3}. With the invalidation of Dennard's scaling principle, the mismatch between Gordon Moore's prediction and manufacturing processes, and the power wall problem, the continuous trend of improving processor efficiency faced a sharp decline due to numerous problems, such as increased leakage currents, which were overcome by applying alternative architectures and the use of multi-core processors for several years \cite{ref3, ref4}. One of the main solutions proposed by researchers to overcome the aforementioned problems has been the use of emerging technologies such as Carbon Nanotube Field-Effect Transistors (CNFETs), Quantum-dot Cellular Automata (QCA), and memristors \cite{ref4, ref5, ref6, ref7}. In addition to emerging technologies, numerous computational methods have been proposed across various scientific communities to replace conventional methods. One of the most promising of these methods is Computation In-Memory (CIM). In the conventional von Neumann architecture, the main memory in the memory hierarchy is located far from the central processor. The significant difference between processor speed and memory bandwidth, combined with the massive volume of data transferred for processing and computation in this architecture, has become a major bottleneck in the design, known as the von Neumann (memory) bottleneck \cite{ref8}. The problem of data transfer in modern processing applications is so serious that, according to estimates from companies such as Google, from 63\% to 90\% of total processing energy consumption is spent solely on data movement, without performing any actual computation \cite{ref9}. The use of CIM architectures that leverage emerging memristive technologies, which enable simultaneous data storage and logic/arithmetic operations, is a promising approach to overcoming the problems posed by the memory wall. Several methods have been proposed for designing arithmetic and logic circuits using memristors, due to their favorable characteristics. In general, these methods can be classified into two categories: stateful and non-stateful \cite{ref10}. Stateful methods are methods that can be used to store and process data completely within the memristive array \cite{ref2, ref4, ref10}. This means that stateful methods such as material implication (IMPLY) \cite{ref11} and Memristor Aided loGIC (MAGIC) \cite{ref12} are memristive processing methods that can be used in the CIM-Array (CIM-A) architecture to significantly reduce data transfer volume \cite{ref13}.

Maintaining information security and privacy across various processing applications related to IoT and AI, where large volumes of data are generated, collected, and transmitted by edge devices, has become a major concern \cite{ref1}. The use of cryptographic mechanisms in these areas is essential. 
%In addition to their widespread application in conventional processing and computing, memristors have also been extensively utilized in random number and hash function generation [], chaotic circuits [], Physical Unclonable Functions (PUFs) [], and cryptographic mechanisms [].
Memristors can be an efficient option for implementing cryptographic platforms due to their small area and power consumption, their resistance to side-channel attacks, and their potential integration into CIM-A architectures to eliminate redundant data transfers \cite{ref1, ref2, ref13, ref22}. Excessive data transfer between the processor and main memory in conventional cryptographic units, which is required to generate and store ciphertext, also increases the risk of attack; this vulnerability can be minimized by applying a CIM-A architecture \cite{ref1}. In \cite{ref17, ref18, ref19}, several methods for generating random and pseudorandom numbers using memristors have been investigated. Implementation and acceleration of hash functions using memristors for processing within and near the memory array have also been discussed in \cite{ref15}. In addition to generating random numbers and hash functions, designers have also considered methods for implementing cryptographic mechanisms that specifically utilize memristors. In \cite{ref1}, the characteristics of the process variations during transistor manufacturing in a single chip are applied to generate random numbers. The non-stateful method of One Transistor-One Resistor (1T-1R) is also employed to XOR the input value stored in the memristors with the aforementioned random number to generate the ciphertext \cite{ref1}. In 2024, the lightweight block cipher GIFT, based on memristors, was used to generate ciphertext on edge devices \cite{ref22}. In this study, the XOR structure is located in the peripheral circuits of the memristive crossbar array (sense amplifiers) \cite{ref22}. Hence, the architecture considered by the authors was categorized as a CIM-Periphery (CIM-P) architecture \cite{ref13}.

Based on the results presented in \cite{ref22, ref23, ref24, ref25, ref26} and an analysis of various lightweight ciphers, it can be concluded that lightweight stream ciphers are more suitable for implementation in devices with limited hardware resources. Using stateful methods in the design of memristor-based circuits not only significantly reduces data transfer and the risk of unauthorized data access during inter-unit transfers, but is also an efficient way to improve circuit metrics. Considering the limitations imposed by power and memory walls, the importance of applying cryptographic mechanisms in inter-unit communications, and the need to minimize data transfer between processing units and memory on a chip, the authors propose the design of IMPLY-based encryption/decryption units for integration into CIM-A architectures, ensuring high compatibility with conventional memristive crossbar arrays. The lightweight stream ciphers evaluated in this article were selected based on global evaluations and standards \cite{ref24, ref25, ref27, ref28}. The selected ciphers have been redesigned and improved based on a reliable stateful method, IMPLY \cite{ref29}, to align seamlessly with the structure of conventional memristive crossbar arrays. Furthermore, the functionality of the proposed designs has been evaluated behaviorally in a steganography application. The results confirmed the practical applicability of the proposed designs in this domain. The main contributions of this paper are as follows:

\begin{enumerate}
	\item Presenting a systematic methodology for porting the classic implementation algorithms of the Trivium \cite{ref27, ref30} and Grain-128a \cite{ref31} lightweight stream ciphers to the structure of CIM-A architectures based on the stateful IMPLY design method;
	\item Developing an algorithm for reducing the computational steps and energy consumption of Linear Feedback Shift Registers (LFSRs) and Nonlinear Feedback Shift Registers (NFSRs) used in the Trivium and Grain-128a lightweight stream ciphers;
	\item Redesigning and evaluating the functional correctness of all the basic logic blocks required in the IMPLY-based improved implementation algorithms of the Trivium and Grain-128a lightweight stream ciphers; and
	\item Examining the functionality and image quality metrics of the proposed memristive encryption and decryption units in a steganography application.
\end{enumerate}

The remainder of this paper is organized into four sections. In Section \ref{sec2}, the basic concepts, definitions, and related work are reviewed.  In Section \ref{sec3}, the proposed algorithm for implementing LFSRs and NFSRs, as well as the step-by-step implementation method for redesigning the Trivium and Grain-128a lightweight stream ciphers based on the IMPLY design method for CIM, are presented. The circuit-level and application-level simulation results are reported in Section \ref{sec4}. Finally, the paper's conclusion is provided in Section \ref{sec5}.

\section{Background and related work} \label{sec2}
\subsection{Lightweight stream ciphers} \label{sec21}
Symmetric ciphers are divided into two categories: stream ciphers and block ciphers \cite{ref27}. In stream ciphers, the message is encrypted by adding it (modulo-2 addition) to a pseudorandom keystream by applying XOR operations. Stream ciphers are often faster than block ciphers and require fewer hardware resources, making them suitable for resource-constrained devices \cite{ref27, ref32}. In general, block ciphers are more susceptible to error propagation when noise corrupts data, and stream ciphers are recommended when the message size is unknown, or the messages are transmitted in continuous streams \cite{ref32}. Stream ciphers were developed based on the One-Time Pad (OTP) cipher, but because of the OTP's limitations, they are used as an alternative \cite{ref23}. Pseudorandom Number Generators (PRNGs) are a main pillar of stream ciphers, which use various structures, including LFSRs. The LFSR structure, in combination with nonlinear logic gates, acquires nonlinear characteristics and is applied in stream ciphers \cite{ref27}.

Over the last three decades, numerous symmetric ciphers have been proposed by various companies and research institutions and evaluated across aspects such as information security, complexity, and the resources required for evaluation \cite{ref23, ref24, ref25, ref26, ref27, ref28, ref31, ref32}. Among the most important of these processes are the eSTREAM project and the evaluations of the US National Institute of Standards and Technology (NIST) \cite{ref23, ref24, ref25, ref26, ref27, ref28, ref31, ref32}. Several stream ciphers have also been examined in these processes. Cryptography is widely applied in processing environments such as the IoT, but many devices in these and similar domains are embedded devices with limited resources \cite{ref24, ref26, ref32}. Therefore, relying on conventional cryptographic algorithms with high computational complexity in these applications is associated with significant limitations. Lightweight cryptography is a different approach from conventional algorithms that enables the maintenance of confidentiality and data integrity on devices with limited resources. Most efficient lightweight ciphers are symmetric ciphers, and in addition to evaluating their security, assessing their circuit metrics, such as energy consumption and resource utilization, is of great importance. A complete introduction and review of the main features of lightweight stream and block ciphers, along with comprehensive comparisons for selecting the appropriate lightweight cipher according to different design constraints, have been reported in \cite{ref23, ref24, ref25, ref26, ref27, ref32}. Based on the evaluations conducted and the results reported in the eSTREAM project and NIST evaluation processes, and considering the structures and characteristics of various lightweight stream ciphers, Trivium \cite{ref30} and Grain-128a \cite{ref31} are selected to be described in detail in the following subsections.

\subsubsection{Trivium} \label{sec211}
\begin{figure}[b]
	\centering
	\includegraphics[scale=0.225]{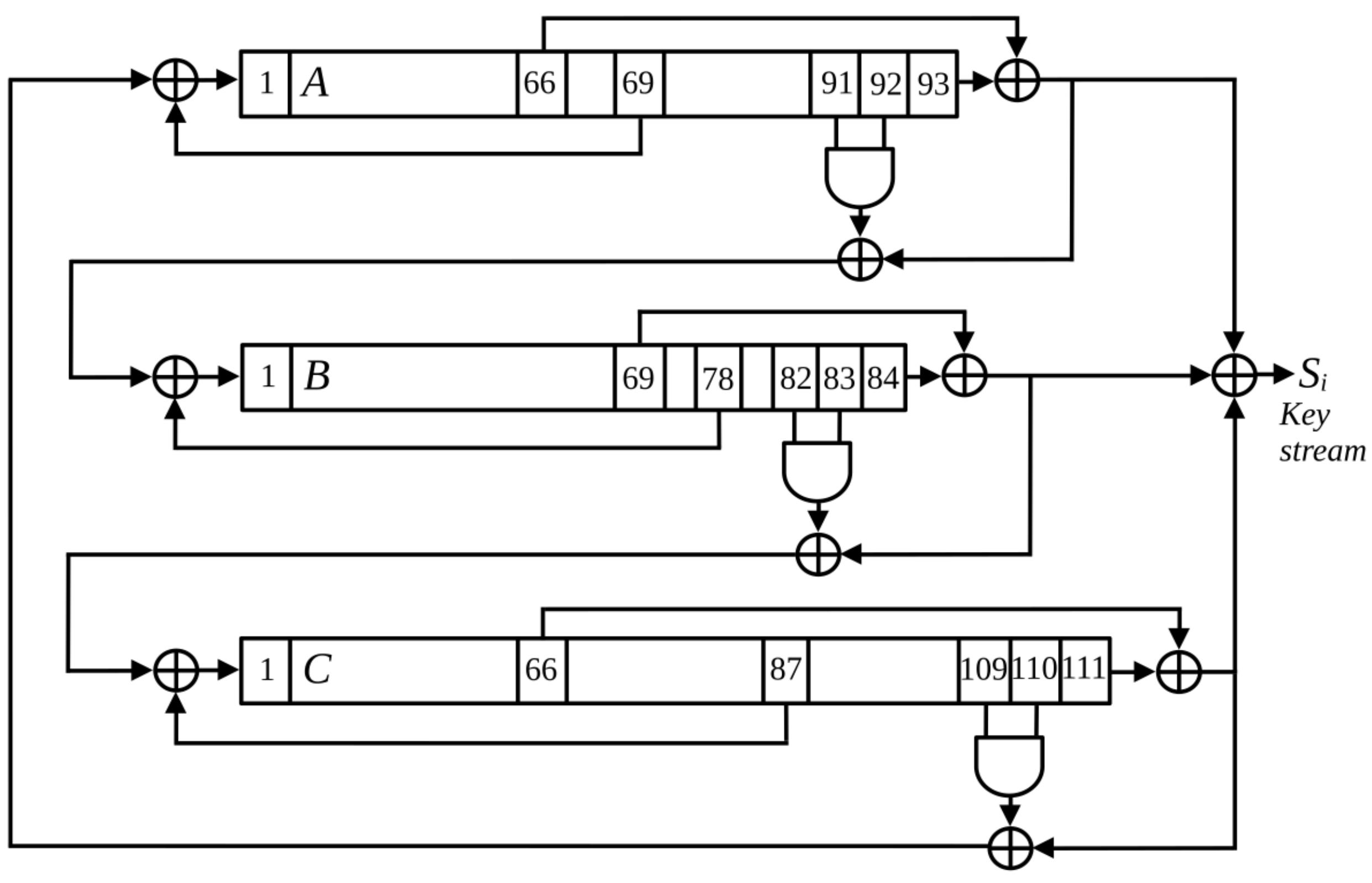}
	\caption{Architecture of the stream cipher Trivium \cite{ref27, ref30}.}
	\label{fig1}
\end{figure}

Trivium is one of the selected lightweight stream ciphers of the eSTREAM project and is also a standardized cipher (ISO/IEC 29192-3:2012) \cite{ref24, ref25}. As shown in Figure \ref{fig1}, the structure of this cipher consists of three shift registers, a 93-bit register ($A$), an 84-bit register ($B$), and a 111-bit register ($C$), combined with nonlinear logic gates \cite{ref27}. By calculating the modulo-2 addition of these three shift registers’ outputs, the keystream ($s_{i}$) is produced \cite{ref27}. The logical equations defining the structure of this lightweight stream cipher are expressed in (\ref{eq1})-(\ref{eq4}). The 80-bit key and Initialization Vector (IV) are placed in the 80 Least Significant Bits (LSBs) of registers $A$ and $B$, respectively, and all other bits of the registers are set to logical ‘0’, except for bits 109-111 of register $C$, which are set to a logical ‘1’ \cite{ref27}. In the initialization phase, the cipher’s structure is clocked 1152 times, and the keystream output from the $1153^{rd}$ cycle onward is used to produce the ciphertext \cite{ref27}.

\begin{equation}
a_i \equiv c_{i-66} + c_{i-111} + c_{i-110} \cdot c_{i-109} + a_{i-69}
\label{eq1}
\end{equation}

\begin{equation}
b_i \equiv a_{i-66} + a_{i-93} + a_{i-92} \cdot a_{i-91} + b_{i-78}
\label{eq2}
\end{equation}

\begin{equation}
c_i \equiv b_{i-69} + b_{i-84} + b_{i-83} \cdot b_{i-82} + c_{i-87}
\label{eq3}
\end{equation}

\begin{equation}
s_i \equiv a_{i-66} + a_{i-93} + b_{i-69} + b_{i-84} + c_{i-66} + c_{i-111}
\label{eq4}
\end{equation}

\subsubsection{Grain-128a} \label{sec212}
\begin{figure}[b]
\centering
\subcaptionbox{}{ \includegraphics[width=0.425\textwidth]{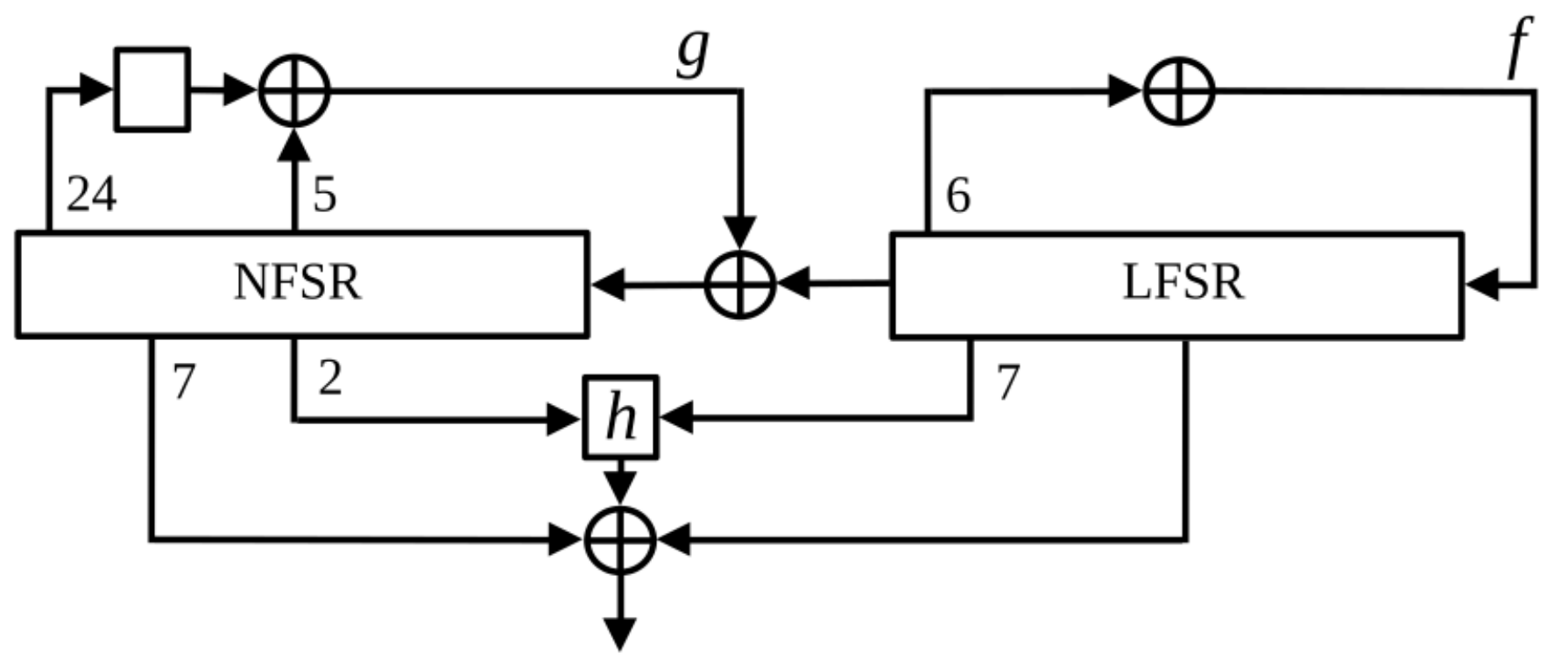} }
\subcaptionbox{}{ \includegraphics[width=0.425\textwidth]{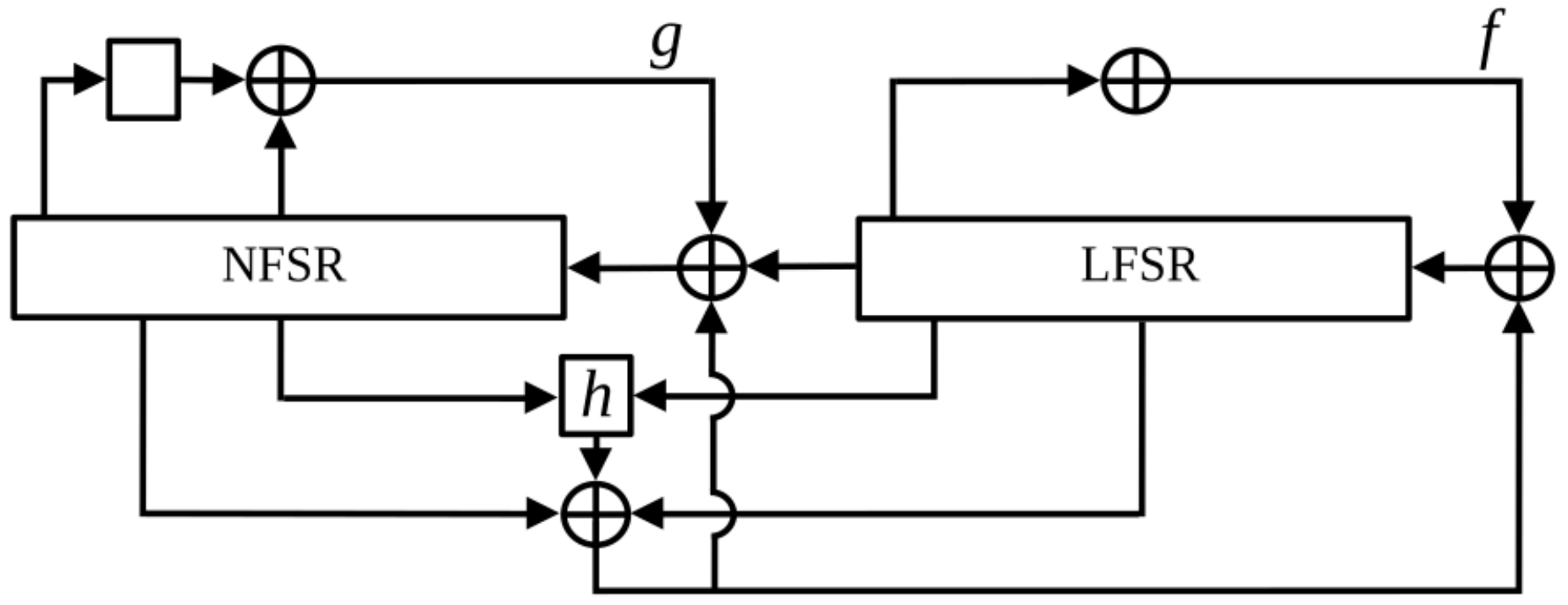} }
\caption{Architecture of Grain-128a: (a) the pre-output generator, and (b) the key initiaization \cite{ref31}.}
\label{fig2}
\end{figure}

Another selected cipher of the eSTREAM project, which has been standardized in ISO/IEC 29167-13:2015 as one of the applicable ciphers for Radio Frequency Identification (RFID) systems, is Grain-128a \cite{ref28}. This cipher can perform authentication in addition to producing a keystream for data encryption/decryption. Here, only the Grain-128a output generation process is discussed. The structure of this cipher is illustrated in Figure \ref{fig2} for the pre-initialization and output generation stages. The main components in the structure of this cipher are the shift registers, each of which is 128 bits. (\ref{eq5}) and (\ref{eq6}) describe the update functions of these two shift registers, respectively \cite{ref31}. The Boolean functions h(x) and y (the output generation function), are also computed according to (\ref{eq7}) and (\ref{eq8}), respectively. The 128-bit key and the 96-bit IV are placed in the NFSR and LFSR, respectively \cite{ref31}. The other 32 bits of the LFSR are initialized from bits 96 to 126 with logical ‘1’, and the last bit (the $127^{th}$ bit) with logical ‘0’ \cite{ref31}. The pre-initialization phase occupies 256 cycles, and from the $257^{th}$ cycle onward, keystream generation begins. In each cycle after the $256^{th}$ one, the modulo-2 addition (XOR) of the keystream and the input data produces the ciphertext \cite{ref31}.

\begin{equation}
s_{i+128} = s_i + s_{i+7} + s_{i+38} + s_{i+70} + s_{i+81} + s_{i+96}
\label{eq5}
\end{equation}

\begin{multline}
b_{i+128} = s_i + b_i + b_{i+26} + b_{i+56} + b_{i+91} + b_{i+96} 
+ b_{i+3} \cdot b_{i+67} + b_{i+11} \cdot b_{i+13} + b_{i+17} \cdot b_{i+18}
+ b_{i+27} \cdot b_{i+59} +  b_{i+40} \\ \cdot b_{i+48}
+ b_{i+61} \cdot b_{i+65} + b_{i+68} \cdot b_{i+84}
+ b_{i+22} \cdot  b_{i+24} \cdot b_{i+25} + b_{i+70} \cdot b_{i+78}
+ b_{i+82} \cdot b_{i+88} + b_{i+92} \cdot b_{i+93} \cdot b_{i+95}
\label{eq6}
\end{multline}

\begin{equation}
h(x) = b_{i+12} \cdot s_{i+8} + s_{i+13} \cdot s_{i+20} + b_{i+95} \cdot s_{i+42}
       + s_{i+60} \cdot s_{i+79} + b_{i+12} \cdot b_{i+95} \cdot s_{i+94}
\label{eq7}
\end{equation}

\begin{equation}
y_i = h(x) + s_{i+93} + \sum_{j \in \mathcal{A}} b_{i+j},
\quad \mathcal{A} \in \{2, 15, 36, 45, 64, 73, 89\}
\label{eq8}
\end{equation}

\subsection{An introduction to memristor} \label{sec22}
\begin{figure}[t]
\centering
\subcaptionbox{}{ \includegraphics[width=0.175\textwidth]{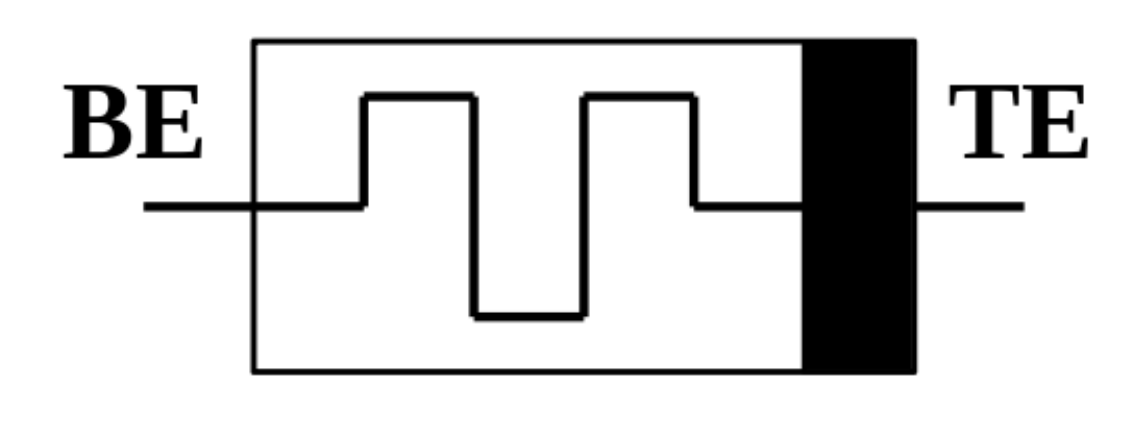} }
\subcaptionbox{}{ \includegraphics[width=0.175\textwidth]{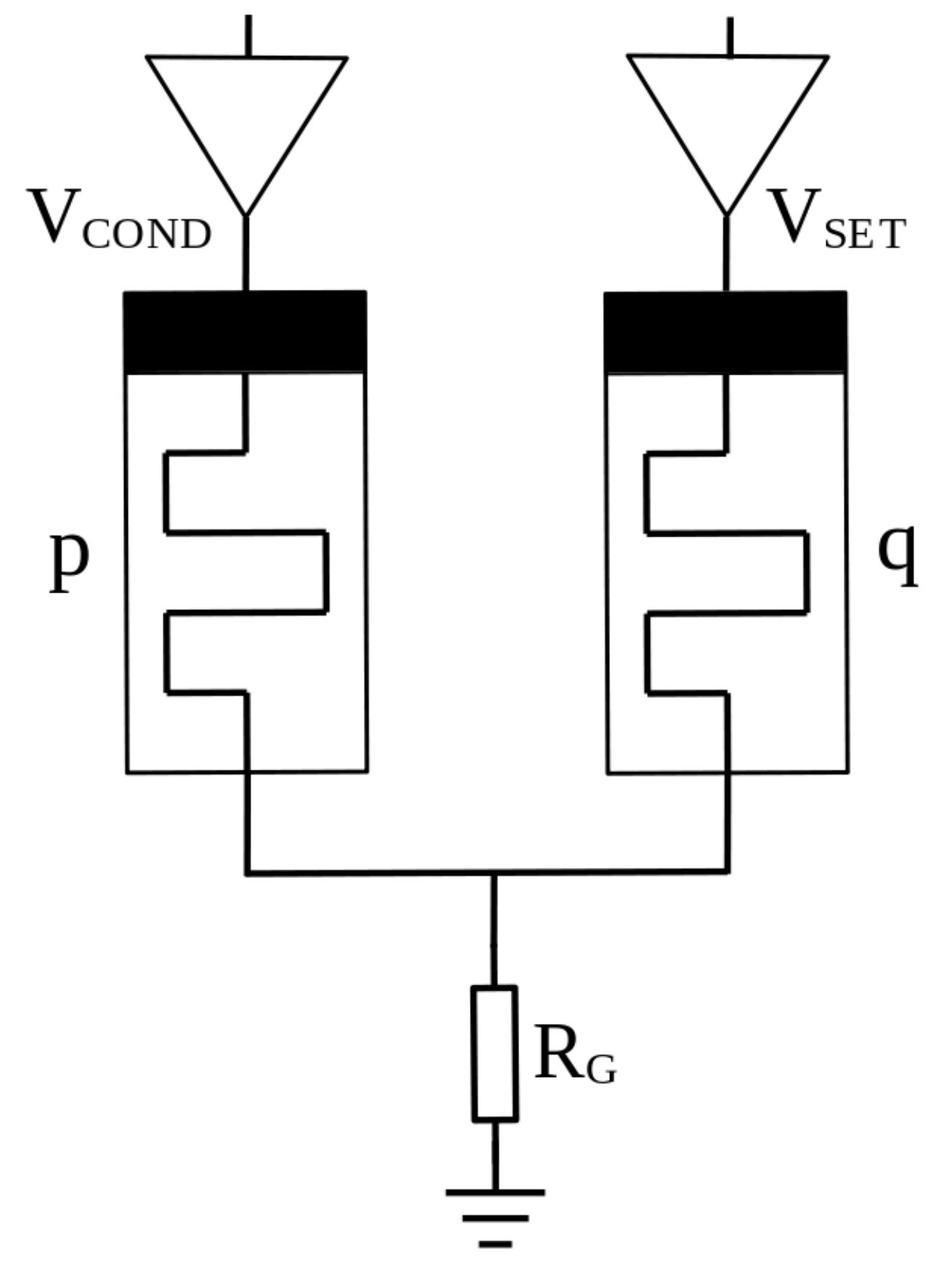} }
\caption{(a) Memristor symbol, and (b) memristive IMPLY gate \cite{ref4}.}
\label{fig3}
\end{figure}

Capacitors, inductors, and resistors are three basic electrical elements that define the relationship between electric charge and voltage, magnetic flux and current, and voltage and current, respectively. Professor L. Chua of the University of California at Berkeley introduced the memristor in the 1970s as the fourth basic electrical element, defining the relationship between electric charge and magnetic flux \cite{ref10}. Four decades later, a team at HP Laboratories led by Professor S. Williams implemented the first physical memristor \cite{ref10}. The memristor, whose symbol is illustrated in Figure \ref{fig3}(a), can act as a non-volatile memory cell, and its resistance determines its logic state. If the resistance of the memristor is at its minimum value ($R_{ON}$), it is generally equivalent to Boolean logic ‘1’, and if the resistance is at its maximum value ($R_{OFF}$), it represents Boolean logic ‘0’ \cite{ref4, ref10}. As shown in Figure \ref{fig3}(a), if a voltage/current is applied from the Top Electrode (TE) of the memristor, its resistance decreases, and if applied from the other terminal, its resistance increases \cite{ref10}. The resistance of the memristor remains unchanged for years when no voltage/current is applied to its terminals, making it an emerging non-volatile memory cell. Memristive CIM has another fundamental aspect: the execution of logic and arithmetic operations. Researchers have proposed several methods for implementing logic and arithmetic functions using memristors, which are generally classified into two categories: stateful and non-stateful \cite{ref10}. In stateful circuit design methods, the resistance of the memristor determines the logic state at the input, output, and intermediate nodes, whereas in non-stateful methods, the logic state is mainly determined by voltage \cite{ref10}. The applicability of stateful design methods in memristive CIM-A has been widely evaluated, and several methods have been developed. Among the most important stateful methods are IMPLY \cite{ref11}, MAGIC \cite{ref12}, and fast and energy-efficient logic in memory (FELIX) \cite{ref33}. Each of these methods has several advantages and disadvantages, and, considering the comparative summary in \cite{ref10, ref29, ref33} and the reliability factor, the IMPLY method has been selected as the stateful method applied in this paper.

\begin{figure}[!]
	\centering
	\includegraphics[scale=0.25]{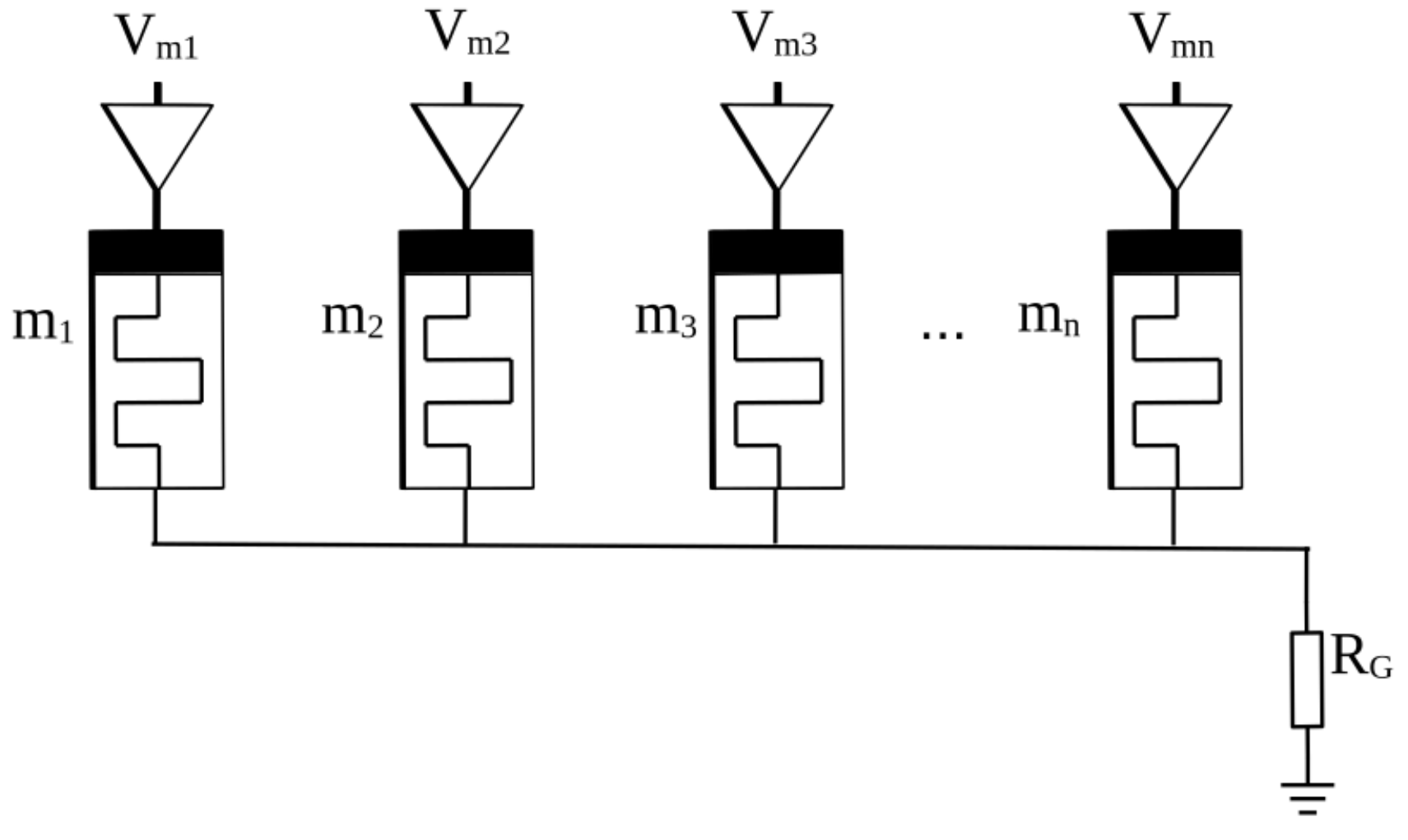}
	\caption{The serial architecture of IMPLY-based logical and arithmetic circuits \cite{ref2, ref4, ref6}.}
	\label{fig4}
\end{figure}

The IMPLY gate architecture and its truth table are illustrated in Figure \ref{fig3}(b) and Table \ref{tab1}, respectively. The IMPLY gate is a universal gate; this means that any logic function can be computed by applying it multiple times in addition to the FALSE (zero) function. For example, the OR logic function can be implemented using two input memristors and one work memristor in two IMPLY-based computational steps, in addition to executing the FALSE function once, as shown in (\ref{eq9}) \cite{ref6}. Work memristors are used alongside input memristors to implement logic and arithmetic functions in the IMPLY method. Storing temporary variables during the execution of IMPLY-based logic and arithmetic functions is the main role of work memristors, although these memristors can also be used to store the outputs. To perform the IMPLY operation, two voltage, $V_{COND}$ and $V_{SET}$, are simultaneously applied to memrsitors `$p$' and `$q$' in Figure \ref{fig3}(b), and the result of ``$p IMP q$" is stored in the memristor `$q$' \cite{ref6}. For the correct execution of the IMPLY operation, two conditions, 1) $V_{COND}<V_{c}<V_{SET}$ ($V_{c}$: the threshold voltage of a memristor) and 2) $R_{ON}<<R_{G}<<R_{OFF}$, must be met \cite{ref4, ref6}. Four architectures, serial, parallel, semi-serial, and semi-parallel, have been introduced for the development of logic and arithmetic circuits using the IMPLY method \cite{ref4, ref6}. The serial architecture is the most compatible with the structure of memristive crossbar arrays; hence, it is an acceptable structure for use in CIM-A \cite{ref4, ref6}. In the serial architecture, the memristors are placed in a row or column of the crossbar array. This architecture is illustrated in Figure \ref{fig4}. The serial architecture has the least hardware complexity among the others, but it can perform only one IMPLY operation per computational step \cite{ref4}. Numerous logic and arithmetic circuits, such as different types of adders and multipliers, have been implemented in the serial architecture using the IMPLY design method \cite{ref2, ref4, ref6, ref11, ref34, ref35, ref36}. Table \ref{tab2} lists the implementation algorithms of serial IMPLY-based basic logic gates and their characteristics.

\begin{equation}
p~OR~q \equiv (p~IMP~0)~IMP~q \equiv (p \to 0)\to q
\label{eq9}
\end{equation}

\begin{table}[t]
	\centering
	\caption{The truth table of an universal IMPLY logic gate \cite{ref4}.}
	\scalebox{0.825}{
	\begin{tabular}{|c|c|c|}
		\hline
		$p$ & $q$ & $p~IMPLY~q$ $\equiv$ $p~IMP~q$ $\equiv$ $p \to q$ \\ \hline
		0 & 0 & 1 \\ \hline
		0 & 1 & 1 \\ \hline
		1 & 0 & 0 \\ \hline
		1 & 1 & 1 \\ \hline 
	\end{tabular}}
	\label{tab1}
\end{table}

\begin{table}[t]
\centering
\caption{Implementation of basic Boolean logic gates using IMPLY method's primitive functions (IMPLY and FALSE) \cite{ref6}}
\scalebox{0.825}{
\begin{tabular}{|c|c|}
\hline
Boolean logic gate & Equivalent IMPLY logic  \\ \hline
NOT p & $p \to 0$ \\ \hline
p OR q & $(p \to 0) \to q$ \\ \hline
p NOR q & $((p \to 0) \to q) \to 0$ \\ \hline 
p NAND q & $p \to (q \to 0)$ \\ \hline 
p AND q & $(p \to (q \to 0)) \to 0$ \\ \hline  
\end{tabular}}
\label{tab2}
\end{table}

\subsection{Memristive cryptography} \label{sec23}
Memristors can be applied in cryptography due to their electrical properties. This subsection briefly discusses a few of these applications.

In \cite{ref1}, an architecture is introduced that generates random numbers based on variations in the threshold slope of Metal Oxide Semiconductor FETs (MOSFETs) in a 1T-1R crossbar array, driven by manufacturing process variations. According to the architecture designed in \cite{ref1}, the threshold slope value is computed by dedicated extraction circuits and stored in a register. Then, this data is read and converted into a specific voltage \cite{ref1}. By applying a non-stateful method introduced in \cite{ref1}, an XOR operation is performed between the data stored in the memristor and the aforementioned random voltage value, and the output ciphertext is generated.

Another application of memristors, widely used in cryptography, is the generation of random numbers. In \cite{ref18, ref19}, different Ring Oscillators (ROs) are designed using varying numbers of transistors and memristors to generate random numbers. In the RO structure introduced in \cite{ref18}, a memristor is placed between the inverters of each row of the RO implemented based on Complementary MOS (CMOS) technology. If the propagation delays of two different rows are not the same, the generated output is logic ‘1’, and otherwise, the output is logic ‘0’ \cite{ref18}. The role of memristors in the RO structure proposed in \cite{ref37} is to replace the PMOS transistors of the inverter’s pull-up network. The use of memristors in the RO structure increases the randomness of this circuit compared to an RO designed entirely with transistors \cite{ref37}. In \cite{ref19}, two rows of single-bit ROs have been used to generate random numbers, where each row consists of multiple CMOS inverters, each connected to a single memristor. The random number generator proposed in \cite{ref19} can be implemented in serial and parallel architectures, so that, if throughput needs to be increased, more data can be processed in the post-processing stage using a parallel-to-serial converter \cite{ref19}. In the post-processing stage, the Trivium stream cipher is used, and after 1152 cycles, a random output bit is generated in each cycle \cite{ref19}. 

The chaotic behavior of the memristor in \cite{ref17} is used to generate noise. The output of the first stage of the proposed structure in \cite{ref17} is a floating-point number. In the second stage, first, the output of the previous stage is multiplied by a fixed number (e.g., $x_{i}$), then the result $y_{i}=x_{i}~mod~64$ is calculated and converted to binary format \cite{ref17}. In the next phase of the second stage, five XOR gates are applied to compute the number of ones in $y_{i}$. Then, the raw bits of the second stage are applied to the Trivium unit in the third stage, and random outputs are generated bit-by-bit \cite{ref17}. In the final stage of the proposed structure in \cite{ref17}, the random outputs are evaluated using the standard NIST statistical tests.

Lightweight encryption based on the GIFT block cipher has been applied within the structure of a memristive crossbar array for healthcare applications in \cite{ref22}. The main idea in \cite{ref22} is to integrate the compact units of the GIFT cipher into the structure of the crossbar array. The assumption in \cite{ref22} is that the generation and exchange of the encryption key between the transmitter and the receiver are handled by methods implemented in previous works. The XOR gate is one of the basic units of the GIFT cipher, which is implemented in \cite{ref22} using Scouting Logic and Dual Sense Amplifier crossbar logic (DSA). This enhances parallel computation and is compatible with other parts of the GIFT cipher \cite{ref22}. These two approaches are among the methods integrated into the CIM-P architecture.

The LFSR is one of the basic structures for generating random numbers. A 4-bit LFSR implementation algorithm using the IMPLY method within a serial architecture has been developed in \cite{ref38}. In the cited paper, the authors designed an algorithm to implement a D-type flip-flop using the IMPLY method, which appears redundant because a memristor can inherently store data like a D flip-flop. However, in \cite{ref38}, an algorithm for implementing a 4-bit LFSR using four D flip-flops and an XOR gate is proposed, employing eight memristors. The IMPLY-based LFSR computes the results in 55 computational steps \cite{ref38}.

\section{Implementation of IMPLY-based Trivium and Grain-128a lightweight stream ciphers} \label{sec3}
In this section, we first examine how to redesign and implement the standard stream ciphers (Trivium and Grain-128a) by applying the serial IMPLY design method for the CIM-A architecture. Then, by applying an innovative method and redesigning the shift register structure, the number of computational steps for both selected stream ciphers is reduced compared to conventional designs, thereby directly reducing energy consumption.

\subsection{IMPLY-based logic gates required for the implementation of Trivium and Grain-128a lightweight stream ciphers} \label{sec31}
In Figures \ref{fig1} and \ref{fig2} and (\ref{eq1})-(\ref{eq8}), the architecture and logic equations required for the implementation of Trivium and Grain-128a ciphers can be seen. In the Trivium’s structure, there are nine two-input XOR gates, one three-input XOR gate, and three two-input AND gates. Forty seven two-input XOR gates, 18 two-input AND gates, five three-input AND gates, and two four-input AND gates are the logic blocks that are employed in the Grain-128a cipher.

To implement logic gates using the IMPLY method, it should be noted that in some cases it is possible that the input data of the logic gate is overwritten and the original value is deleted, so, in some cases, it is necessary to design and use an algorithm that preserves the value of the input data in the input memristors in these gates. For example, the algorithm introduced in \cite{ref6} for implementing a two-input XOR gate incur a loss of one of the inputs. Therefore, to implement a two-input XOR gate, two algorithms are used, as shown in Tables \ref{tab3} and \ref{tab4}. The first algorithm computes the output in nine computational steps using two work memristors, and one of the input memristors loses its original value (see (\ref{eq10})). The second implementation algorithm obtains the output using three work memristors in eleven computational steps. In the second algorithm, the logic values of the the input memristors are preserved.

\begin{table}[t]
\centering
\caption{Implementation algorithm of a destructive IMPLY-based serial XOR logic gate introduced in \cite{ref6}}
\scalebox{0.825}{
\begin{tabular}{|c|c|c|}
\hline
\textbf{Step} & \textbf{Operation} & \textbf{Equivalent Logic} \\ \hline
\textbf{1} & $s_{1}=0$ & FALSE($s_{1}$) \\ \hline
\textbf{2} & $s_{2}=0$ & FALSE($s_{2}$) \\ \hline
\textbf{3} & $a \to s_{1}= s_{1}^{'}$ & $NOT(a)$ \\ \hline
\textbf{4} & $b \to s_{2}= s_{2}^{'}$ & $NOT(b)$ \\ \hline
\textbf{5} & $s_{1}^{'} \to s_{2}^{'}=s_{2}^{''}$ & $NOT(a) \to NOT(b)$ \\ \hline
\textbf{6} & $s_{1}=0$ & FALSE($s_{1}$) \\ \hline
\textbf{7} & $s_{2}^{''} \to s_{1}=s_{1}^{'}$ & $NOT(NOT(a) \to NOT(b))$  \\ \hline
\textbf{8} & $a \to b=b^{'}$ & $a \to b$ \\ \hline
\textbf{9} & $b^{'} \to s_{1}^{'}=s_{1}^{''}$ & $XOR(a,~b)$ \\ \hline
\end{tabular}}
\label{tab3}
\end{table}

\begin{equation}
p~XOR~q \equiv (p \to q) \to ((q \to p) \to 0)
\label{eq10}
\end{equation}

\begin{table}[t]
\centering
\caption{Implementation algorithm of a non-destructive IMPLY-based serial XOR logic gate}
\scalebox{0.825}{
\begin{tabular}{|c|c|c|}
\hline
\textbf{Step} & \textbf{Operation} & \textbf{Equivalent Logic} \\ \hline
\textbf{1} & $s_{1}=0$ & FALSE($s_{1}$) \\ \hline
\textbf{2} & $s_{2}=0$ & FALSE($s_{2}$) \\ \hline
\textbf{3} & $s_{3}=0$ & FALSE($s_{3}$) \\ \hline
\textbf{4} & $a \to s_{1}= s_{1}^{'}$ & $NOT(a)$ \\ \hline
\textbf{5} & $b \to s_{2}= s_{2}^{'}$ & $NOT(b)$ \\ \hline
\textbf{6} & $s_{2}^{'} \to s_{3}= s_{3}^{'}$ & $b$ \\ \hline
\textbf{7} & $s_{1}^{'} \to s_{2}^{'}=s_{2}^{''}$ & $NOT(a) \to NOT(b)$ \\ \hline
\textbf{8} & $a \to s_{3}^{'}=s_{3}^{''} $ & $a \to b$ \\ \hline
\textbf{9} & $s_{1}=0$ & FALSE($s_{1}$) \\ \hline
\textbf{10} & $s_{2}^{''} \to s_{1}=s_{1}^{'}$ & $NOT(NOT(a) \to NOT(b))$  \\ \hline
\textbf{11} & $s_{3}^{''} \to s_{1}^{'}=s_{1}^{''}$ & $XOR(a,~b)$ \\ \hline
\end{tabular}}
\label{tab4}
\end{table}

Two methods can be used to implement XOR gates with three inputs and beyond. If the designer's goal is to minimize the computational steps and input data preservation is not important, it is recommended to use the algorithm introduced in \cite{ref6}, where the XOR result of two inputs is first calculated in nine steps, then in each of the nine steps after that, the XOR result of the previous inputs is XORed with the next input and the final result is calculated. The second method, however, focuses on preserving the input data. In this method, the first two inputs are first XORed using three work memristors according to the algorithm presented in Table \ref{tab4}, and the result is placed in a work memristor. After that, other input is XORed with the result of the previous step using the algorithm written in Table \ref{tab3}.

In addition to XOR logic gates, two-input, three-input, and four-input AND gates are required to design the two proposed lightweight ciphers. The two-input AND gate can be implemented in a serial architecture using the IMPLY method, with two work memristors and in five computational steps, as shown in Table \ref{tab2}. For the three-input AND gate, the third input can be ANDed with the result of the AND of the first two inputs, and the result can be stored in one of the work memristors in the final step (the $10^{th}$ computational step). By examining (\ref{eq11}) and Table \ref{tab5}, it can be concluded that by overlapping the computational steps of two two-input AND gates, the number of computational steps of a three-input AND gate can be reduced to six by executing two consecutive AND algorithms. The four-input AND gate can also be implemented in 11 computational steps by executing the algorithms for the three-input AND gate and the two-input AND gate, according to Table \ref{tab6}.

\begin{equation}
AND(a,~b,~c) \equiv (c \to [((b \to (a \to)) \to 0) \to 0]) \to 0 \equiv (c \to (b \to (a \to 0))) \to 0
\label{eq11}
\end{equation}

\begin{table}[b]
\centering
\caption{Implementation algorithm of a three-input IMPLY-based serial AND gate}
\scalebox{0.825}{
\begin{tabular}{|c|c|c|}
\hline
\textbf{Step} & \textbf{Operation} & \textbf{Equivalent Logic} \\ \hline
\textbf{1} & $s_{1}=0$ & FALSE($s_{1}$) \\ \hline
\textbf{2} & $s_{2}=0$ & FALSE($s_{2}$) \\ \hline
\textbf{3} & $a \to s_{1}= s_{1}^{'}$ & $NOT(a)$ \\ \hline
\textbf{4} & $b \to s_{1}^{'}= s_{1}^{''}$ & $b \to NOT(a)$ \\ \hline
\textbf{5} & $c \to s_{1}^{''}= s_{1}^{'''}$ & $c \to (b \to NOT(a))$ \\ \hline
\textbf{6} & $s_{1}^{'''} \to s_{2}=s_{2}^{'}$ & $AND(a,~b,~c)$ \\ \hline
\end{tabular}}
\label{tab5}
\end{table}

\begin{table}[t]
\centering
\caption{Implementation algorithm of a four-input IMPLY-based serial AND gate}
\scalebox{0.825}{
\begin{tabular}{|c|c|c|}
\hline
\textbf{Step} & \textbf{Operation} & \textbf{Equivalent Logic} \\ \hline
\textbf{1} & $s_{1}=0$ & FALSE($s_{1}$) \\ \hline
\textbf{2} & $s_{2}=0$ & FALSE($s_{2}$) \\ \hline
\textbf{3} & $a \to s_{1}= s_{1}^{'}$ & $NOT(a)$ \\ \hline
\textbf{4} & $b \to s_{1}^{'}= s_{1}^{''}$ & $b \to NOT(a)$ \\ \hline
\textbf{5} & $c \to s_{1}^{''}= s_{1}^{'''}$ & $c \to (b \to NOT(a))$ \\ \hline
\textbf{6} & $s_{1}^{'''} \to s_{2}=s_{2}^{'}$ & $AND(a,~b,~c)$ \\ \hline
\textbf{7} & $s_{1}=0$ & FALSE($s_{1}$) \\ \hline
\textbf{8} & $d \to s_{1}= s_{1}^{'}$ & $NOT(d)$ \\ \hline
\textbf{9} & $s_{2}^{'} \to s_{1}^{'}= s_{1}^{''}$ & $AND(a,~b,~c) \to NOT(d)$ \\ \hline
\textbf{10} & $s_{2}=0$ & FALSE($s_{2}$) \\ \hline
\textbf{11} & $s_{1}^{''} \to s_{2}=s_{2}^{'}$ & $AND(a,~b,~c,~d)$ \\ \hline
\end{tabular}}
\label{tab6}
\end{table}

\subsection{Proposed method for implementing shift registers required by lightweight stream ciphers}
In conventional architectures, it is common to use D flip-flops for implementing shift registers. In the CIM-A architecture, as designers implement different logical and arithmetic circuits with memristors, there is no need to design or use memristive D flip-flops. The purpose of using memristors in the CIM-A architecture is the same as that of a D flip-flop, as memristors are inherently capable of storing data within the crossbar array. Memristors can also be used for data processing. In a shift register, the stored data is shifted bit-by-bit in each cycle from the $m^{th}$ cell to the $m+1^{th}$ or $m-1^{th}$ cell, depending on the shift direction. In the LSB or MSB of each register, a new bit is entered in each cycle, computed using a special or random calculation method, or assigned a fixed value. Therefore, in each cycle, a new bit is placed in the register, and one is removed. In a D flip-flop, data is placed on the data input line before the rising edge of the clock pulse, and it is stored after the rising edge. This process is completely different in the IMPLY-based serial architecture considered here. In this architecture, a buffer should be applied to shift each bit from one cell to another. The buffer implementation algorithm (two consecutive inverters, see Table \ref{tab2}) shifts data in four computational steps using two work memristors.

Using a buffer to shift data in the shift registers of both the Trivium and Grain-128a ciphers is a reliable method. The disadvantage of this method is that each bit of data is shifted in four computational steps. From Figures \ref{fig1} and \ref{fig2}, a key point for efficiently designing IMPLY-based shift registers can be identified. In each cycle of processing data in these two ciphers, a few cells in each register are directly involved in the computation, and a large number of bits are only shifted from one cell to its neighbor at the end of each cycle. According to Figure \ref{fig1}, only five bits (bits 66, 69, and 91-93) of Trivium’s shift register $A$ are used in the computations of the two-input XOR, two-input AND, and three-input XOR logic gates. The other bits of this register (in the Trivium cipher structure) are only shifted one position to the right (the more significant bit) at the end of each cycle. In this process, it is important that the aforementioned bits have the correct value in each processing cycle to maintain the security of the cipher. The same principle applies to the other two registers of the Trivium cipher and the two registers of the Grain-128a cipher. Considering the above points, the number of computational steps can be significantly reduced by replacing some buffers with inverters, so that each buffer is replaced by an inverter, reducing the number of computational steps by two.

The use of the aforementioned method is accompanied by certain design constraints. Specifically, the correctness of the bits used in the computations must be ensured in every processing cycle. Therefore, an algorithm must be proposed and applied that maintains the integrity of the data stored in the memristors required for logical and arithmetic processing in all cycles. Hence, in addition to improving circuit evaluation criteria such as the number of computational steps, the cipher operates correctly without any functional disruption. The proposed algorithm is as follows:

The data located in consecutive memristors and used directly in logical processing (such as two-input XOR gates or two-input AND gates) must be transferred from the main memristor (the first memristor, e.g., $k$) to its neighbor (the second memristor, e.g., $k^{'}$) in each cycle using only a buffer. In other words, a work memristor and a target memristor (the second memristor, $k^{'}$), where the data is stored, are used to transfer the bit of the first memristor to the target memristor in four computational steps using the IMPLY-based buffer implementation algorithm. This process is performed in all processing cycles. For example, consider Figure \ref{fig1}. In shift register $A$, bits 91 ($k$) and 92 ($k^{'}$) are both involved in the computations. In each of the 1152 initialization cycles (in the warm-up phase) or from cycle 1153 onwards, only the buffer implementation algorithm must be used to transfer data from the $91^{st}$ bit of shift register $A$ to its $92^{nd}$ bit. The main problem, however, concerns the first memristor ($k$). If the distance of the input data from the first memristor ($k$) is an odd number of memristors (for example, a distance of one, three, five, etc.), the desired value is shifted with the buffer in the cycle preceding the data shift. Here, the input data is either the logical value stored in the memristor used in logical and arithmetic computations during the previous cycle, or the input bit of the register. For example, in the Trivium’s first shift register ($A$), the input memristor for the $91^{st}$ cell ($k$) in the proposed algorithm is the output of the $66^{th}$ memristor ($k^{''}$), and for the $65^{th}$ cell ($k^{'''}$), the input of the register is considered as the input. The distance of the data from the target memristor is of great importance. If the distance is equal to ``$2 \times n-1$" bits and `$n$' buffers and `$n$' inverters are used, then only an inverter is used in the data shifts of cycle ``$2 \times n+1$" onwards. For example, if the distance between the input data and the target memristor is nine memristors, `$n$' is equal to five, meaning that the combination of buffers and inverters (respectively) is used for 10 consecutive cycles, and from the $11^{th}$ cycle onwards, only the inverter is used. If the number of memristors placed between the input data and target memristor is ``$2\times n$", the buffers and inverters are used `$n$' times, and from the ``$2 \times n+1^{th}$" cycle onwards, only the buffer is applied to shift data. If the distance between the input data and the target memristor is 10 memristors, the buffers and inverters are used five times, and only the buffer is used to shift data from the $11^{th}$ cycle onwards. The other bits stored in the memristors before the first memristor ($k$) and after the second memristor ($k^{'}$) are shifted only by the inverters (see the inverter implementation algorithm in Table \ref{tab2}), and the shifted values are stored in the adjacent memristors. If the input data is directly stored in the first memristor, only the buffer should be used. For a better understanding of the proposed algorithm, please refer to Tables \ref{tab7} and \ref{tab8}.

In these toy examples, the conventional and proposed shift register implementation algorithms are shown over 12 cycles. In Tables \ref{tab7} and \ref{tab8}, `$n$' equals 5 and 6, respectively. In these tables, the columns marked in green refer to the memristors involved in the computations ($k$ and $k^{'}$).

\begin{table}[t]
\caption{A toy example of an 8-bit conventional (buffer-based) and an 8-bit proposed (inverter-based) shift register with a distance of 5 cells}
\label{tab7}
\centering
\resizebox{0.825\textwidth}{!}{%
\begin{tabular}{|cccccccccc|l|cccccccccc|}
\cline{1-10} \cline{12-21}
\multicolumn{10}{|c|}{Conventional buffer-based 8-bit shift register} &  & \multicolumn{10}{c|}{Proposed inverter-based 8-bit shift register} \\ \cline{1-10} \cline{12-21} 
\multicolumn{1}{|c|}{Cycle} & \multicolumn{1}{c|}{Input} & \multicolumn{1}{c|}{0} & \multicolumn{1}{c|}{1} & \multicolumn{1}{c|}{2} & \multicolumn{1}{c|}{3} & \multicolumn{1}{c|}{4} & \multicolumn{1}{c|}{\cellcolor[HTML]{83E019}5} & \multicolumn{1}{c|}{\cellcolor[HTML]{83E019}6} & 7 &  & \multicolumn{1}{c|}{Cycle} & \multicolumn{1}{c|}{Input} & \multicolumn{1}{c|}{0} & \multicolumn{1}{c|}{1} & \multicolumn{1}{c|}{2} & \multicolumn{1}{c|}{3} & \multicolumn{1}{c|}{4} & \multicolumn{1}{c|}{\cellcolor[HTML]{83E019}5} & \multicolumn{1}{c|}{\cellcolor[HTML]{83E019}6} & 7 \\ \cline{1-10} \cline{12-21} 
\multicolumn{1}{|c|}{0} & \multicolumn{1}{c|}{1} & \multicolumn{1}{c|}{1} & \multicolumn{1}{c|}{0} & \multicolumn{1}{c|}{1} & \multicolumn{1}{c|}{1} & \multicolumn{1}{c|}{1} & \multicolumn{1}{c|}{\cellcolor[HTML]{83E019}0} & \multicolumn{1}{c|}{\cellcolor[HTML]{83E019}1} & 0 &  & \multicolumn{1}{c|}{0} & \multicolumn{1}{c|}{1} & \multicolumn{1}{c|}{1} & \multicolumn{1}{c|}{0} & \multicolumn{1}{c|}{1} & \multicolumn{1}{c|}{1} & \multicolumn{1}{c|}{1} & \multicolumn{1}{c|}{\cellcolor[HTML]{83E019}0} & \multicolumn{1}{c|}{\cellcolor[HTML]{83E019}1} & 0 \\ \cline{1-10} \cline{12-21} 
\multicolumn{1}{|c|}{1} & \multicolumn{1}{c|}{0} & \multicolumn{1}{c|}{1} & \multicolumn{1}{c|}{1} & \multicolumn{1}{c|}{0} & \multicolumn{1}{c|}{1} & \multicolumn{1}{c|}{1} & \multicolumn{1}{c|}{\cellcolor[HTML]{83E019}1} & \multicolumn{1}{c|}{\cellcolor[HTML]{83E019}0} & 1 &  & \multicolumn{1}{c|}{1} & \multicolumn{1}{c|}{0} & \multicolumn{1}{c|}{0} & \multicolumn{1}{c|}{0} & \multicolumn{1}{c|}{1} & \multicolumn{1}{c|}{0} & \multicolumn{1}{c|}{0} & \multicolumn{1}{c|}{\cellcolor[HTML]{83E019}1} & \multicolumn{1}{c|}{\cellcolor[HTML]{83E019}0} & 0 \\ \cline{1-10} \cline{12-21} 
\multicolumn{1}{|c|}{2} & \multicolumn{1}{c|}{1} & \multicolumn{1}{c|}{0} & \multicolumn{1}{c|}{1} & \multicolumn{1}{c|}{1} & \multicolumn{1}{c|}{0} & \multicolumn{1}{c|}{1} & \multicolumn{1}{c|}{\cellcolor[HTML]{83E019}1} & \multicolumn{1}{c|}{\cellcolor[HTML]{83E019}1} & 0 &  & \multicolumn{1}{c|}{2} & \multicolumn{1}{c|}{1} & \multicolumn{1}{c|}{1} & \multicolumn{1}{c|}{1} & \multicolumn{1}{c|}{1} & \multicolumn{1}{c|}{0} & \multicolumn{1}{c|}{1} & \multicolumn{1}{c|}{\cellcolor[HTML]{83E019}1} & \multicolumn{1}{c|}{\cellcolor[HTML]{83E019}1} & 1 \\ \cline{1-10} \cline{12-21} 
\multicolumn{1}{|c|}{3} & \multicolumn{1}{c|}{0} & \multicolumn{1}{c|}{1} & \multicolumn{1}{c|}{0} & \multicolumn{1}{c|}{1} & \multicolumn{1}{c|}{1} & \multicolumn{1}{c|}{0} & \multicolumn{1}{c|}{\cellcolor[HTML]{83E019}1} & \multicolumn{1}{c|}{\cellcolor[HTML]{83E019}1} & 1 &  & \multicolumn{1}{c|}{3} & \multicolumn{1}{c|}{0} & \multicolumn{1}{c|}{0} & \multicolumn{1}{c|}{0} & \multicolumn{1}{c|}{0} & \multicolumn{1}{c|}{0} & \multicolumn{1}{c|}{1} & \multicolumn{1}{c|}{\cellcolor[HTML]{83E019}1} & \multicolumn{1}{c|}{\cellcolor[HTML]{83E019}1} & 0 \\ \cline{1-10} \cline{12-21} 
\multicolumn{1}{|c|}{4} & \multicolumn{1}{c|}{0} & \multicolumn{1}{c|}{0} & \multicolumn{1}{c|}{1} & \multicolumn{1}{c|}{0} & \multicolumn{1}{c|}{1} & \multicolumn{1}{c|}{1} & \multicolumn{1}{c|}{\cellcolor[HTML]{83E019}0} & \multicolumn{1}{c|}{\cellcolor[HTML]{83E019}1} & 1 &  & \multicolumn{1}{c|}{4} & \multicolumn{1}{c|}{0} & \multicolumn{1}{c|}{1} & \multicolumn{1}{c|}{1} & \multicolumn{1}{c|}{1} & \multicolumn{1}{c|}{1} & \multicolumn{1}{c|}{1} & \multicolumn{1}{c|}{\cellcolor[HTML]{83E019}0} & \multicolumn{1}{c|}{\cellcolor[HTML]{83E019}1} & 0 \\ \cline{1-10} \cline{12-21} 
\multicolumn{1}{|c|}{5} & \multicolumn{1}{c|}{1} & \multicolumn{1}{c|}{0} & \multicolumn{1}{c|}{0} & \multicolumn{1}{c|}{1} & \multicolumn{1}{c|}{0} & \multicolumn{1}{c|}{1} & \multicolumn{1}{c|}{\cellcolor[HTML]{83E019}1} & \multicolumn{1}{c|}{\cellcolor[HTML]{83E019}0} & 1 &  & \multicolumn{1}{c|}{5} & \multicolumn{1}{c|}{1} & \multicolumn{1}{c|}{1} & \multicolumn{1}{c|}{0} & \multicolumn{1}{c|}{0} & \multicolumn{1}{c|}{0} & \multicolumn{1}{c|}{0} & \multicolumn{1}{c|}{\cellcolor[HTML]{83E019}1} & \multicolumn{1}{c|}{\cellcolor[HTML]{83E019}0} & 0 \\ \cline{1-10} \cline{12-21} 
\multicolumn{1}{|c|}{6} & \multicolumn{1}{c|}{1} & \multicolumn{1}{c|}{1} & \multicolumn{1}{c|}{0} & \multicolumn{1}{c|}{0} & \multicolumn{1}{c|}{1} & \multicolumn{1}{c|}{0} & \multicolumn{1}{c|}{\cellcolor[HTML]{83E019}1} & \multicolumn{1}{c|}{\cellcolor[HTML]{83E019}1} & 0 &  & \multicolumn{1}{c|}{6} & \multicolumn{1}{c|}{1} & \multicolumn{1}{c|}{0} & \multicolumn{1}{c|}{0} & \multicolumn{1}{c|}{1} & \multicolumn{1}{c|}{1} & \multicolumn{1}{c|}{1} & \multicolumn{1}{c|}{\cellcolor[HTML]{83E019}1} & \multicolumn{1}{c|}{\cellcolor[HTML]{83E019}1} & 1 \\ \cline{1-10} \cline{12-21} 
\multicolumn{1}{|c|}{7} & \multicolumn{1}{c|}{0} & \multicolumn{1}{c|}{1} & \multicolumn{1}{c|}{1} & \multicolumn{1}{c|}{0} & \multicolumn{1}{c|}{0} & \multicolumn{1}{c|}{1} & \multicolumn{1}{c|}{\cellcolor[HTML]{83E019}0} & \multicolumn{1}{c|}{\cellcolor[HTML]{83E019}1} & 1 &  & \multicolumn{1}{c|}{7} & \multicolumn{1}{c|}{0} & \multicolumn{1}{c|}{0} & \multicolumn{1}{c|}{1} & \multicolumn{1}{c|}{1} & \multicolumn{1}{c|}{0} & \multicolumn{1}{c|}{0} & \multicolumn{1}{c|}{\cellcolor[HTML]{83E019}0} & \multicolumn{1}{c|}{\cellcolor[HTML]{83E019}1} & 0 \\ \cline{1-10} \cline{12-21} 
\multicolumn{1}{|c|}{8} & \multicolumn{1}{c|}{1} & \multicolumn{1}{c|}{0} & \multicolumn{1}{c|}{1} & \multicolumn{1}{c|}{1} & \multicolumn{1}{c|}{0} & \multicolumn{1}{c|}{0} & \multicolumn{1}{c|}{\cellcolor[HTML]{83E019}1} & \multicolumn{1}{c|}{\cellcolor[HTML]{83E019}0} & 1 &  & \multicolumn{1}{c|}{8} & \multicolumn{1}{c|}{1} & \multicolumn{1}{c|}{1} & \multicolumn{1}{c|}{1} & \multicolumn{1}{c|}{0} & \multicolumn{1}{c|}{0} & \multicolumn{1}{c|}{1} & \multicolumn{1}{c|}{\cellcolor[HTML]{83E019}1} & \multicolumn{1}{c|}{\cellcolor[HTML]{83E019}0} & 0 \\ \cline{1-10} \cline{12-21} 
\multicolumn{1}{|c|}{9} & \multicolumn{1}{c|}{0} & \multicolumn{1}{c|}{1} & \multicolumn{1}{c|}{0} & \multicolumn{1}{c|}{1} & \multicolumn{1}{c|}{1} & \multicolumn{1}{c|}{0} & \multicolumn{1}{c|}{\cellcolor[HTML]{83E019}0} & \multicolumn{1}{c|}{\cellcolor[HTML]{83E019}1} & 0 &  & \multicolumn{1}{c|}{9} & \multicolumn{1}{c|}{0} & \multicolumn{1}{c|}{0} & \multicolumn{1}{c|}{0} & \multicolumn{1}{c|}{0} & \multicolumn{1}{c|}{1} & \multicolumn{1}{c|}{1} & \multicolumn{1}{c|}{\cellcolor[HTML]{83E019}0} & \multicolumn{1}{c|}{\cellcolor[HTML]{83E019}1} & 1 \\ \cline{1-10} \cline{12-21} 
\multicolumn{1}{|c|}{10} & \multicolumn{1}{c|}{1} & \multicolumn{1}{c|}{0} & \multicolumn{1}{c|}{1} & \multicolumn{1}{c|}{0} & \multicolumn{1}{c|}{1} & \multicolumn{1}{c|}{1} & \multicolumn{1}{c|}{\cellcolor[HTML]{83E019}0} & \multicolumn{1}{c|}{\cellcolor[HTML]{83E019}0} & 1 &  & \multicolumn{1}{c|}{10} & \multicolumn{1}{c|}{1} & \multicolumn{1}{c|}{1} & \multicolumn{1}{c|}{1} & \multicolumn{1}{c|}{1} & \multicolumn{1}{c|}{1} & \multicolumn{1}{c|}{0} & \multicolumn{1}{c|}{\cellcolor[HTML]{83E019}0} & \multicolumn{1}{c|}{\cellcolor[HTML]{83E019}0} & 0 \\ \cline{1-10} \cline{12-21} 
\multicolumn{1}{|c|}{11} & \multicolumn{1}{c|}{1} & \multicolumn{1}{c|}{1} & \multicolumn{1}{c|}{0} & \multicolumn{1}{c|}{1} & \multicolumn{1}{c|}{0} & \multicolumn{1}{c|}{1} & \multicolumn{1}{c|}{\cellcolor[HTML]{83E019}1} & \multicolumn{1}{c|}{\cellcolor[HTML]{83E019}0} & 0 &  & \multicolumn{1}{c|}{11} & \multicolumn{1}{c|}{1} & \multicolumn{1}{c|}{0} & \multicolumn{1}{c|}{0} & \multicolumn{1}{c|}{0} & \multicolumn{1}{c|}{0} & \multicolumn{1}{c|}{0} & \multicolumn{1}{c|}{\cellcolor[HTML]{83E019}1} & \multicolumn{1}{c|}{\cellcolor[HTML]{83E019}0} & 1 \\ \cline{1-10} \cline{12-21} 
\multicolumn{1}{|c|}{12} & \multicolumn{1}{c|}{} & \multicolumn{1}{c|}{1} & \multicolumn{1}{c|}{1} & \multicolumn{1}{c|}{0} & \multicolumn{1}{c|}{1} & \multicolumn{1}{c|}{0} & \multicolumn{1}{c|}{\cellcolor[HTML]{83E019}1} & \multicolumn{1}{c|}{\cellcolor[HTML]{83E019}1} & 0 &  & \multicolumn{1}{c|}{12} & \multicolumn{1}{c|}{} & \multicolumn{1}{c|}{0} & \multicolumn{1}{c|}{1} & \multicolumn{1}{c|}{1} & \multicolumn{1}{c|}{1} & \multicolumn{1}{c|}{1} & \multicolumn{1}{c|}{\cellcolor[HTML]{83E019}1} & \multicolumn{1}{c|}{\cellcolor[HTML]{83E019}1} & 1 \\ \cline{1-10} \cline{12-21} 
\end{tabular}%
}
\end{table}

\begin{table}[t]
\caption{A toy example of an 8-bit conventional (buffer-based) and an 8-bit proposed (inverter-based) shift register with a distance of 6 cells}
\label{tab8}
\centering
\resizebox{0.825\textwidth}{!}{%
\begin{tabular}{|cccccccccc|l|cccccccccc|}
\cline{1-10} \cline{12-21}
\multicolumn{10}{|c|}{Conventional buffer-based 8-bit shift register} &  & \multicolumn{10}{c|}{Proposed inverter-based 8-bit shift register} \\ \cline{1-10} \cline{12-21} 
\multicolumn{1}{|c|}{Cycle} & \multicolumn{1}{c|}{Input} & \multicolumn{1}{c|}{0} & \multicolumn{1}{c|}{1} & \multicolumn{1}{c|}{2} & \multicolumn{1}{c|}{3} & \multicolumn{1}{c|}{4} & \multicolumn{1}{c|}{5} & \multicolumn{1}{c|}{\cellcolor[HTML]{83E019}6} & \cellcolor[HTML]{83E019}7 &  & \multicolumn{1}{c|}{Cycle} & \multicolumn{1}{c|}{Input} & \multicolumn{1}{c|}{0} & \multicolumn{1}{c|}{1} & \multicolumn{1}{c|}{2} & \multicolumn{1}{c|}{3} & \multicolumn{1}{c|}{4} & \multicolumn{1}{c|}{5} & \multicolumn{1}{c|}{\cellcolor[HTML]{83E019}6} & \cellcolor[HTML]{83E019}7 \\ \cline{1-10} \cline{12-21} 
\multicolumn{1}{|c|}{0} & \multicolumn{1}{c|}{1} & \multicolumn{1}{c|}{1} & \multicolumn{1}{c|}{0} & \multicolumn{1}{c|}{1} & \multicolumn{1}{c|}{1} & \multicolumn{1}{c|}{1} & \multicolumn{1}{c|}{0} & \multicolumn{1}{c|}{\cellcolor[HTML]{83E019}1} & \cellcolor[HTML]{83E019}0 &  & \multicolumn{1}{c|}{0} & \multicolumn{1}{c|}{1} & \multicolumn{1}{c|}{1} & \multicolumn{1}{c|}{0} & \multicolumn{1}{c|}{1} & \multicolumn{1}{c|}{1} & \multicolumn{1}{c|}{1} & \multicolumn{1}{c|}{0} & \multicolumn{1}{c|}{\cellcolor[HTML]{83E019}1} & \cellcolor[HTML]{83E019}0 \\ \cline{1-10} \cline{12-21} 
\multicolumn{1}{|c|}{1} & \multicolumn{1}{c|}{0} & \multicolumn{1}{c|}{1} & \multicolumn{1}{c|}{1} & \multicolumn{1}{c|}{0} & \multicolumn{1}{c|}{1} & \multicolumn{1}{c|}{1} & \multicolumn{1}{c|}{1} & \multicolumn{1}{c|}{\cellcolor[HTML]{83E019}0} & \cellcolor[HTML]{83E019}1 &  & \multicolumn{1}{c|}{1} & \multicolumn{1}{c|}{0} & \multicolumn{1}{c|}{0} & \multicolumn{1}{c|}{0} & \multicolumn{1}{c|}{1} & \multicolumn{1}{c|}{0} & \multicolumn{1}{c|}{0} & \multicolumn{1}{c|}{0} & \multicolumn{1}{c|}{\cellcolor[HTML]{83E019}0} & \cellcolor[HTML]{83E019}1 \\ \cline{1-10} \cline{12-21} 
\multicolumn{1}{|c|}{2} & \multicolumn{1}{c|}{1} & \multicolumn{1}{c|}{0} & \multicolumn{1}{c|}{1} & \multicolumn{1}{c|}{1} & \multicolumn{1}{c|}{0} & \multicolumn{1}{c|}{1} & \multicolumn{1}{c|}{1} & \multicolumn{1}{c|}{\cellcolor[HTML]{83E019}1} & \cellcolor[HTML]{83E019}0 &  & \multicolumn{1}{c|}{2} & \multicolumn{1}{c|}{1} & \multicolumn{1}{c|}{1} & \multicolumn{1}{c|}{1} & \multicolumn{1}{c|}{1} & \multicolumn{1}{c|}{0} & \multicolumn{1}{c|}{1} & \multicolumn{1}{c|}{1} & \multicolumn{1}{c|}{\cellcolor[HTML]{83E019}1} & \cellcolor[HTML]{83E019}0 \\ \cline{1-10} \cline{12-21} 
\multicolumn{1}{|c|}{3} & \multicolumn{1}{c|}{0} & \multicolumn{1}{c|}{1} & \multicolumn{1}{c|}{0} & \multicolumn{1}{c|}{1} & \multicolumn{1}{c|}{1} & \multicolumn{1}{c|}{0} & \multicolumn{1}{c|}{1} & \multicolumn{1}{c|}{\cellcolor[HTML]{83E019}1} & \cellcolor[HTML]{83E019}1 &  & \multicolumn{1}{c|}{3} & \multicolumn{1}{c|}{0} & \multicolumn{1}{c|}{0} & \multicolumn{1}{c|}{0} & \multicolumn{1}{c|}{0} & \multicolumn{1}{c|}{0} & \multicolumn{1}{c|}{1} & \multicolumn{1}{c|}{0} & \multicolumn{1}{c|}{\cellcolor[HTML]{83E019}1} & \cellcolor[HTML]{83E019}1 \\ \cline{1-10} \cline{12-21} 
\multicolumn{1}{|c|}{4} & \multicolumn{1}{c|}{0} & \multicolumn{1}{c|}{0} & \multicolumn{1}{c|}{1} & \multicolumn{1}{c|}{0} & \multicolumn{1}{c|}{1} & \multicolumn{1}{c|}{1} & \multicolumn{1}{c|}{0} & \multicolumn{1}{c|}{\cellcolor[HTML]{83E019}1} & \cellcolor[HTML]{83E019}1 &  & \multicolumn{1}{c|}{4} & \multicolumn{1}{c|}{0} & \multicolumn{1}{c|}{1} & \multicolumn{1}{c|}{1} & \multicolumn{1}{c|}{1} & \multicolumn{1}{c|}{1} & \multicolumn{1}{c|}{1} & \multicolumn{1}{c|}{0} & \multicolumn{1}{c|}{\cellcolor[HTML]{83E019}1} & \cellcolor[HTML]{83E019}1 \\ \cline{1-10} \cline{12-21} 
\multicolumn{1}{|c|}{5} & \multicolumn{1}{c|}{1} & \multicolumn{1}{c|}{0} & \multicolumn{1}{c|}{0} & \multicolumn{1}{c|}{1} & \multicolumn{1}{c|}{0} & \multicolumn{1}{c|}{1} & \multicolumn{1}{c|}{1} & \multicolumn{1}{c|}{\cellcolor[HTML]{83E019}0} & \cellcolor[HTML]{83E019}1 &  & \multicolumn{1}{c|}{5} & \multicolumn{1}{c|}{1} & \multicolumn{1}{c|}{1} & \multicolumn{1}{c|}{0} & \multicolumn{1}{c|}{0} & \multicolumn{1}{c|}{0} & \multicolumn{1}{c|}{0} & \multicolumn{1}{c|}{0} & \multicolumn{1}{c|}{\cellcolor[HTML]{83E019}0} & \cellcolor[HTML]{83E019}1 \\ \cline{1-10} \cline{12-21} 
\multicolumn{1}{|c|}{6} & \multicolumn{1}{c|}{1} & \multicolumn{1}{c|}{1} & \multicolumn{1}{c|}{0} & \multicolumn{1}{c|}{0} & \multicolumn{1}{c|}{1} & \multicolumn{1}{c|}{0} & \multicolumn{1}{c|}{1} & \multicolumn{1}{c|}{\cellcolor[HTML]{83E019}1} & \cellcolor[HTML]{83E019}0 &  & \multicolumn{1}{c|}{6} & \multicolumn{1}{c|}{1} & \multicolumn{1}{c|}{0} & \multicolumn{1}{c|}{0} & \multicolumn{1}{c|}{1} & \multicolumn{1}{c|}{1} & \multicolumn{1}{c|}{1} & \multicolumn{1}{c|}{1} & \multicolumn{1}{c|}{\cellcolor[HTML]{83E019}1} & \cellcolor[HTML]{83E019}0 \\ \cline{1-10} \cline{12-21} 
\multicolumn{1}{|c|}{7} & \multicolumn{1}{c|}{0} & \multicolumn{1}{c|}{1} & \multicolumn{1}{c|}{1} & \multicolumn{1}{c|}{0} & \multicolumn{1}{c|}{0} & \multicolumn{1}{c|}{1} & \multicolumn{1}{c|}{0} & \multicolumn{1}{c|}{\cellcolor[HTML]{83E019}1} & \cellcolor[HTML]{83E019}1 &  & \multicolumn{1}{c|}{7} & \multicolumn{1}{c|}{0} & \multicolumn{1}{c|}{0} & \multicolumn{1}{c|}{1} & \multicolumn{1}{c|}{1} & \multicolumn{1}{c|}{0} & \multicolumn{1}{c|}{0} & \multicolumn{1}{c|}{0} & \multicolumn{1}{c|}{\cellcolor[HTML]{83E019}1} & \cellcolor[HTML]{83E019}1 \\ \cline{1-10} \cline{12-21} 
\multicolumn{1}{|c|}{8} & \multicolumn{1}{c|}{1} & \multicolumn{1}{c|}{0} & \multicolumn{1}{c|}{1} & \multicolumn{1}{c|}{1} & \multicolumn{1}{c|}{0} & \multicolumn{1}{c|}{0} & \multicolumn{1}{c|}{1} & \multicolumn{1}{c|}{\cellcolor[HTML]{83E019}0} & \cellcolor[HTML]{83E019}1 &  & \multicolumn{1}{c|}{8} & \multicolumn{1}{c|}{1} & \multicolumn{1}{c|}{1} & \multicolumn{1}{c|}{1} & \multicolumn{1}{c|}{0} & \multicolumn{1}{c|}{0} & \multicolumn{1}{c|}{1} & \multicolumn{1}{c|}{1} & \multicolumn{1}{c|}{\cellcolor[HTML]{83E019}0} & \cellcolor[HTML]{83E019}1 \\ \cline{1-10} \cline{12-21} 
\multicolumn{1}{|c|}{9} & \multicolumn{1}{c|}{0} & \multicolumn{1}{c|}{1} & \multicolumn{1}{c|}{0} & \multicolumn{1}{c|}{1} & \multicolumn{1}{c|}{1} & \multicolumn{1}{c|}{0} & \multicolumn{1}{c|}{0} & \multicolumn{1}{c|}{\cellcolor[HTML]{83E019}1} & \cellcolor[HTML]{83E019}0 &  & \multicolumn{1}{c|}{9} & \multicolumn{1}{c|}{0} & \multicolumn{1}{c|}{0} & \multicolumn{1}{c|}{0} & \multicolumn{1}{c|}{0} & \multicolumn{1}{c|}{1} & \multicolumn{1}{c|}{1} & \multicolumn{1}{c|}{0} & \multicolumn{1}{c|}{\cellcolor[HTML]{83E019}1} & \cellcolor[HTML]{83E019}0 \\ \cline{1-10} \cline{12-21} 
\multicolumn{1}{|c|}{10} & \multicolumn{1}{c|}{1} & \multicolumn{1}{c|}{0} & \multicolumn{1}{c|}{1} & \multicolumn{1}{c|}{0} & \multicolumn{1}{c|}{1} & \multicolumn{1}{c|}{1} & \multicolumn{1}{c|}{0} & \multicolumn{1}{c|}{\cellcolor[HTML]{83E019}0} & \cellcolor[HTML]{83E019}1 &  & \multicolumn{1}{c|}{10} & \multicolumn{1}{c|}{1} & \multicolumn{1}{c|}{1} & \multicolumn{1}{c|}{1} & \multicolumn{1}{c|}{1} & \multicolumn{1}{c|}{1} & \multicolumn{1}{c|}{0} & \multicolumn{1}{c|}{0} & \multicolumn{1}{c|}{\cellcolor[HTML]{83E019}0} & \cellcolor[HTML]{83E019}1 \\ \cline{1-10} \cline{12-21} 
\multicolumn{1}{|c|}{11} & \multicolumn{1}{c|}{1} & \multicolumn{1}{c|}{1} & \multicolumn{1}{c|}{0} & \multicolumn{1}{c|}{1} & \multicolumn{1}{c|}{0} & \multicolumn{1}{c|}{1} & \multicolumn{1}{c|}{1} & \multicolumn{1}{c|}{\cellcolor[HTML]{83E019}0} & \cellcolor[HTML]{83E019}0 &  & \multicolumn{1}{c|}{11} & \multicolumn{1}{c|}{1} & \multicolumn{1}{c|}{0} & \multicolumn{1}{c|}{0} & \multicolumn{1}{c|}{0} & \multicolumn{1}{c|}{0} & \multicolumn{1}{c|}{0} & \multicolumn{1}{c|}{1} & \multicolumn{1}{c|}{\cellcolor[HTML]{83E019}0} & \cellcolor[HTML]{83E019}0 \\ \cline{1-10} \cline{12-21} 
\multicolumn{1}{|c|}{12} & \multicolumn{1}{c|}{} & \multicolumn{1}{c|}{1} & \multicolumn{1}{c|}{1} & \multicolumn{1}{c|}{0} & \multicolumn{1}{c|}{1} & \multicolumn{1}{c|}{0} & \multicolumn{1}{c|}{1} & \multicolumn{1}{c|}{\cellcolor[HTML]{83E019}1} & \cellcolor[HTML]{83E019}0 &  & \multicolumn{1}{c|}{12} & \multicolumn{1}{c|}{} & \multicolumn{1}{c|}{0} & \multicolumn{1}{c|}{1} & \multicolumn{1}{c|}{1} & \multicolumn{1}{c|}{1} & \multicolumn{1}{c|}{1} & \multicolumn{1}{c|}{1} & \multicolumn{1}{c|}{\cellcolor[HTML]{83E019}1} & \cellcolor[HTML]{83E019}0 \\ \cline{1-10} \cline{12-21} 
\end{tabular}%
}
\end{table}

\subsection{Step-by-step implementation of the Trivium and Grain-128a lightweight stream ciphers applying the IMPLY method in a serial architecture} \label{sec33}
In this subsection, after examining the implementation details of the basic logical and arithmetic blocks from the previous subsections, the implementation of each IMPLY-based lightweight stream cipher, the Trivium and Grain-128a, is analyzed.

\subsubsection{Step-by-step implementation algorithm of the IMPLY-based Trivium cipher} \label{sec331}
The implementation of this cipher is divided into three parts. The first part deals with the initialization of the shift registers using random numbers, which, as in \cite{ref22}, is assumed by default in this research. The second part is the Trivium initialization phase, where data is calculated and processed 1152 times, and the third part starts at cycle 1153, where one bit of the keystream is generated in each cycle. To store data and implement this cipher using IMPLY logic, 288 memristors are required for implementing the registers, along with five work memristors $s_{0}$-$s_{4}$ and one output memristor.

In the computations related to register $A$ in the structure shown in Figure \ref{fig1}, two two-input XOR gates and one two-input AND gate are required, yielding a total of 23 computational steps. Three work memristors are used in the computations of this part, and the results are stored in $s_{0}$ and $s_{2}$. At the end of this stage, the logical values of $s_{1}$ and $A_{93}$ are no longer required, so these memristors can be used to store intermediate values in the next stages. The computations for registers $B$ and $C$, similar to the computations for register $A$, are performed using two two-input XOR gates and one two-input AND gate, and the number of computational steps required for each of these registers is also 23 steps. The data required for subsequent processing steps related to registers $B$ and $C$ are stored in memristors $A_{93}$ and $s_{3}$ (for register $B$) and in memristors $B_{84}$ and $s_{4}$ (for register $C$).

After each processing cycle of the Trivium cipher, a two-input XOR gate is used to calculate the input bit for each of the registers ($A_{1}$, $B_{1}$, and $C_{1}$). Hence, three two-input XOR gates are required to compute the values to be stored in the first memristor of each register, requiring a total of 27 computational steps for this stage without the need for any new work memristors. After computing the input bits of the registers, the implementation algorithm for the two-input XOR gate must be executed twice to compute the output $s_{i}$, and this is also completed in 18 computational steps.

The remainder of this subsection explains the details of how data is shifted bit-by-bit in the shift registers of the Trivium lightweight stream cipher. In the previous subsection, the proposed method for implementing fast, energy-efficient shift registers using IMPLY-based inverters and buffers was described in detail. This method is applied in all three shift registers of the Trivium cipher’s structure, as shown in Figure \ref{fig1}. The functionality of the shift register using the proposed method can be examined and observed in detail in Tables \ref{tab9}-\ref{tab11}.

\begin{table}[t]
\centering
\caption{Details of the bit‑by‑bit data shifting process in Trivium's shift register $A$ based on the proposed method}
\scalebox{0.825}{
\begin{tabular}{|c|c|c|}
\hline
\textbf{Memristors involved} & \textbf{Applied} & \textbf{Additional Details} \\
\textbf{in data shifting} & \textbf{circuits} & \textbf{} \\ \hline
\textbf{$A_{92} \to A_{93}$} & 1 buffer & \\ \hline
\textbf{$A_{91} \to A_{92}$} & 1 buffer & \\ \hline
\textbf{$A_{90} \to A_{91}$} & Buffers and inverters & 21 intervening memristors; 11 buffers/inverters, then inverters only.\\ \hline
\textbf{$A_{70} \to A_{90}$} & 20 inverters & \\ \hline
\textbf{$A_{69} \to A_{70}$} & 1 inverter & \\ \hline
\textbf{$A_{68} \to A_{69}$} & Buffers and inverters & 2 intervening memristors; 1 buffers/inverters, then buffers only.\\ \hline
\textbf{$A_{67} \to A_{68}$} & 1 inverter & \\ \hline
\textbf{$A_{66} \to A_{67}$} & 1 inverter & \\ \hline
\textbf{$A_{65} \to A_{66}$} & Buffers and inverters & 65 intervening memristors; 33 buffers/inverters, then inverters only.\\ \hline
\textbf{$A_{1} \to A_{65}$} & 64 inverters & \\ \hline
\textbf{$Input \to A_{1}$} & 1 inverter & \\ \hline
\end{tabular}}
\label{tab9}
\end{table}

\begin{table}[t]
\centering
\caption{Details of the bit‑by‑bit data shifting process in Trivium's shift register $B$ based on the proposed method}
\scalebox{0.825}{
\begin{tabular}{|c|c|c|}
\hline
\textbf{Memristors involved} & \textbf{Applied} & \textbf{Additional Details} \\
\textbf{in data shifting} & \textbf{circuits} & \textbf{} \\ \hline
\textbf{$B_{83} \to B_{84}$} & 1 buffer & \\ \hline
\textbf{$B_{82} \to B_{83}$} & 1 buffer & \\ \hline
\textbf{$B_{81} \to B_{82}$} & Buffers and inverters & 3 intervening memristors; 2 buffers/inverters, then inverters only.\\ \hline
\textbf{$B_{80} \to B_{81}$} & 1 inverter & \\ \hline
\textbf{$B_{79} \to B_{80}$} & 1 inverter & \\ \hline
\textbf{$B_{78} \to B_{79}$} & 1 inverter & \\ \hline
\textbf{$B_{77} \to B_{78}$} & Buffers and inverters & 8 intervening memristors; 4 buffers/inverters, then buffers only.\\ \hline
\textbf{$B_{70} \to B_{77}$} & 7 inverters & \\ \hline
\textbf{$A_{69} \to A_{70}$} & 1 inverter & 2 intervening memristors; 1 buffers/inverters, then buffers only.\\ \hline
\textbf{$B_{68} \to B_{69}$} & Buffers and inverters & 68 intervening memristors; 34 buffers/inverters, then buffers only.\\ \hline
\textbf{$B_{1} \to B_{68}$} & 67 inverters & \\ \hline
\textbf{$Input \to B_{1}$} & 1 inverter & \\ \hline
\end{tabular}}
\label{tab10}
\end{table}

\begin{table}[t]
\centering
\caption{Details of the bit‑by‑bit data shifting process in Trivium's shift register $C$ based on the proposed method}
\scalebox{0.825}{
\begin{tabular}{|c|c|c|}
\hline
\textbf{Memristors involved} & \textbf{Applied} & \textbf{Additional Details} \\
\textbf{in data shifting} & \textbf{circuits} & \textbf{} \\ \hline
\textbf{$C_{110} \to C_{111}$} & 1 buffer & \\ \hline
\textbf{$C_{109} \to C_{110}$} & 1 buffer & \\ \hline
\textbf{$C_{108} \to C_{109}$} & Buffers and inverters & 21 intervening memristors; 11 buffers/inverters, then inverters only.\\ \hline
\textbf{$C_{88} \to C_{108}$} & 20 inverters & \\ \hline
\textbf{$C_{87} \to C_{88}$} & 1 inverter & \\ \hline
\textbf{$C_{86} \to C_{87}$} & Buffers and inverters & 20 intervening memristors; 10 buffers/inverters, then buffers only.\\ \hline
\textbf{$C_{67} \to C_{86}$} & 19 inverter & \\ \hline
\textbf{$C_{66} \to C_{67}$} & 1 inverter & \\ \hline
\textbf{$C_{65} \to C_{66}$} & Buffers and inverters & 65 intervening memristors; 33 buffers/inverters, then inverters only.\\ \hline
\textbf{$C_{1} \to C_{65}$} & 64 inverters & \\ \hline
\textbf{$Input \to C_{1}$} & 1 inverter & \\ \hline
\end{tabular}}
\label{tab11}
\end{table}

As an example, the details of how the stored bits in the memristors of register $A$ are shifted are explained here. Based on the contents of Table \ref{tab9}, it can be concluded that two buffers and 88 inverters must be used to shift data directly in each processing cycle. Special bits are those stored in the memristors before they are involved in logical and arithmetic computations; shifting data in these memristors must be performed carefully. Accordingly, three buffers and three inverters in the first and second cycles, 40 buffers and 20 inverters in the third to $22^{nd}$ cycles, and 66 inverters and 66 buffers in the $23^{rd}$ to $66^{th}$ cycles are required to implement shift register $A$ based on the proposed method, correctly shifting bits to specific memristors. From the $67^{th}$ cycle onwards, two inverters and one buffer are required in each cycle to transfer data in shift register $A$. The proposed method of shifting data in shift register $A$ was examined in detail. A similar procedure is employed in the other two shift registers ($B$ and $C$). Table \ref{tab12} summarizes the number of inverters and buffers for different cycles of each register used in the Trivium lightweight stream cipher.

\begin{table}[t]
\centering
\caption{Inverter and buffer counts for Trivium's shift registers $A$, $B$, and $C$ per cycle}
\scalebox{0.825}{
\begin{tabular}{|c|c|c|c|c|c|c|c|c|}
\hline
\textbf{Cycles} & \textbf{No. of} & \textbf{No. of} &\textbf{Cycles} & \textbf{No. of} & \textbf{No. of} & \textbf{Cycles} & \textbf{No. of} & \textbf{No. of} \\
 & \textbf{buffers} & \textbf{inverters} & & \textbf{buffers} & \textbf{inverters} &  & \textbf{buffers} & \textbf{inverters} \\ \hline
\multicolumn{3}{|c|}{Shift register $A$} & \multicolumn{3}{|c|}{Shift register $B$} & \multicolumn{3}{|c|}{Shift register $C$} \\ \hline
1-2 & 7 & 179 & 1-2 & 7 & 161 & 1-20 & 70 & 2150 \\ \hline
3-22 & 80 & 1780 & 3-4 & 7 & 179 & 21-22 & 8 & 214 \\ \hline
23-66 & 154 & 3938 & 5-8 & 12 & 324 & 23-66 & 154 & 4730 \\ \hline
67 & 3 & 90 & 9-68 & 210 & 4830 & 67 & 3 & 108 \\ \hline
68 and onwards & 3 & 90 & 69 and onwards & 4 & 80 & 68 and onwards & 3 & 108 \\ \hline
\end{tabular}}
\label{tab12}
\end{table}

\subsubsection{Step-by-step implementation algorithm of the IMPLY-based Grain-128a cipher} \label{sec332}
The lightweight Grain-128a stream cipher also performs computations in three stages, like the Trivium cipher. The first stage initializes the shift registers, which will not be discussed in this research. The second stage, called the ``pre-initialization” stage, is performed in 256 processing cycles, and the keystream is generated from the $257^{th}$ round, which is the third stage of implementing this lightweight stream cipher. In this subsection of the article, each part of the second and third stages of the Grain-128a cipher will be examined and implemented based on stateful IMPLY logic according to Figure \ref{fig2}. Two hundred and sixty two memristors are needed to implement this cipher, including two sets of 128 memristors for implementing 128-bit LFSR and 128-bit NFSR, along with six work memristors. If the designer needs to store the output ($y$) separately in a specific memristor, the number of memristors increases to 263 (see Figure \ref{fig2}).

To calculate the output ($y$), two values, $h(x)$ and a set of $b-terms$, must be calculated according to (\ref{eq7}) and (\ref{eq8}). The final result is obtained by performing the two-input XOR algorithm of the values stored in the $93^{rd}$ bit of the LFSR with the two-input XOR product of the calculated values $h(x)$ and $b-terms$. The logical computations of the function $h(x)$ include four two-input AND gates, four two-input XOR gates, and one three-input AND gate, which are performed in 62 computational steps, and the final result is stored in the first work memristor ($s_{0}$). The implementation of the computations related to $b-terms$ is executed using six two-input XOR logic gates, one of which is implemented by applying the algorithm of Table \ref{tab4}, $XOR(b_{2}, b_{15})$, and the other gates are calculated using the algorithm of Table \ref{tab3}. The important point in these computations is that the values stored in the main memristors remain completely intact. The output of these calculations is obtained after 56 computational steps. The output of this step is also stored in a work memristor ($s_{2}$). The final output is also calculated using two two-input XOR gates in 18 computational steps, as shown in (\ref{eq8}). The computations related to the LFSR feedback, performed in the ``pre-initialization” and keystream stages according to (\ref{eq5}), require five two-input XOR logic gates and a total of 47 computational steps. The first XOR is calculated using the algorithm of Table \ref{tab4}, and the other XORs are calculated using the algorithm of Table \ref{tab3} to preserve the input values.

In the second and third stages of the Grain-128a cipher implementation, the NFSR feedback calculation is divided into two parts. It has a higher computational complexity than the other logical parts of this cipher. The first part computes the subsections mentioned in (\ref{eq6}), which involve several XOR and AND gates. The first part specifically includes the computation of 15 two-input XOR gates: 14 are computed using the algorithm in Table \ref{tab3}, and one is computed using the algorithm in Table \ref{tab4}. Eight two-input AND gates (five computational steps) and three three-input AND gates (six computational steps) are also used to calculate (\ref{eq6}). It should be noted that all the main memristors remain intact after these calculations, with 195 computational steps. The final result of this part is also stored in one of the work memristors ($s_{2}$).

In the ``pre-initialization” stage of the Grain-128a lightweight stream cipher, which is the first 256 rounds of its execution, the output of each round (the same value stored in the first work memristor ($s_{0}$)) is XORed as one of the inputs of two two-input XOR gates with the feedbacks of the registers in this cipher, and the results of each XOR are considered as the input bits of each register.

If the data shift in the Grain-128a’s LFSR and NFSR, which consists of 256 bits in total, is performed using IMPLY-based buffers, it takes 1024 computational steps per cycle. By using the proposed method and replacing inverters with buffers in these registers, the computational steps per cycle can be significantly reduced. The implementation of Grain-128a’s shift registers is much more complex than the registers in the Trivium cipher. A similar approach to Tables \ref{tab9}-\ref{tab11} is also applied to determine the details of performing data shifts in the Grain-128a cipher shift registers. According to the proposed approach, 114 inverters and two buffers are required to transfer data from each memristor to its neighbor in the LFSR per cycle. The problem's complexity, however, stems from the other 12 shifts directly involved in the computations. The number of inverters and buffers applied to shift bits that contribute to computations each cycle is tabulated in Table \ref{tab13}. A similar approach should be applied to examine the bit-by-bit shift process in an NFSR to reduce its hardware complexity significantly. In all processing cycles, 92 inverters and 12 buffers are required to shift data between memristors. The shift of the bits stored in the memristors that participate in the logical computations follows the specific characteristics of the proposed algorithm. A summary of the number of buffers and inverters required per cycle for the 24 bits involved in the logical computations of the NFSR is tabulated in Table \ref{tab14}.

\begin{table}[t]
\centering
\caption{Number of buffers and inverters required per cycle for the bits involved in the logical computations of the Grain-128a's LFSR}
\scalebox{0.825}{
\begin{tabular}{|c|c|c|c|c|c|c|c|c|}
\hline
\textbf{Cycles} & \textbf{No. of} & \textbf{No. of} &\textbf{Cycles} & \textbf{No. of} & \textbf{No. of} & \textbf{Cycles} & \textbf{No. of} & \textbf{No. of} \\
 & \textbf{buffers} & \textbf{inverters} & & \textbf{buffers} & \textbf{inverters} &  & \textbf{buffers} & \textbf{inverters} \\ \hline
1-2 & 12 & 12 & 3-4 & 10 & 14 & 5-6 & 10 & 14\\ \hline
7-8 & 12 & 12 & 9-10 & 13 & 11 & 11-12 & 12 & 12\\ \hline
13-18 & 33 & 39 & 19-32 & 63 & 105 & 33 and onwards & 4 & 8\\ \hline
\end{tabular}}
\label{tab13}
\end{table}

\begin{table}[t]
\centering
\caption{Number of buffers and inverters required per cycle for the bits involved in the logical computations of the Grain-128a's NFSR}
\scalebox{0.825}{
\begin{tabular}{|c|c|c|c|c|c|c|c|c|}
\hline
\textbf{Cycles} & \textbf{No. of} & \textbf{No. of} &\textbf{Cycles} & \textbf{No. of} & \textbf{No. of} & \textbf{Cycles} & \textbf{No. of} & \textbf{No. of} \\
 & \textbf{buffers} & \textbf{inverters} & & \textbf{buffers} & \textbf{inverters} &  & \textbf{buffers} & \textbf{inverters} \\ \hline
1-2 & 24 & 24 & 3-4 & 18 & 30 & 5-6 & 16 & 32\\ \hline
7-8 & 16 & 32 & 9-10 & 15 & 33 & 11-12 & 15 & 33\\ \hline
13-32 & 15 & 33 & 33-34 & 16 & 32 & 35 and onwards & 8 & 16\\ \hline
\end{tabular}}
\label{tab14}
\end{table}

\section{Simulation results, discussion, and evaluation} \label{sec4}
In the previous section, the implementation of IMPLY-based Trivium and Grain-128a, two lightweight stream ciphers, in the serial architecture was investigated and reported. In subsection \ref{sec41}, first, the memristor model applied in circuit-level simulation is introduced; second, the method of applying this model to perform the simulation is explained; and in the last part of this subsection, the simulation results are presented and expounded in detail. In the second subsection, the behavioral simulation results for the proposed encryption/decryption structures (based on Trivium and Grain-128a) and their application to image steganography will be reported.

\subsection{Circuit-level simulation, method, and results} \label{sec41}
For simulating memristor-based logical and arithmetic circuits, various behavioral and physical SPICE models have been presented. The logical and arithmetic blocks applied in the structure of Trivium and Grain-128a in this paper have been simulated by applying the modified open source Voltage ThrEshold Adaptive Memristor (VTEAM) model, which has been fitted according to the parameters of a discrete memristor (BS-AF-W model) developed by ``Knowm inc.” \cite{ref2}. This model has been applied in \cite{ref2, ref4, ref6, ref39, ref40, ref41}. The model parameters and standard coefficients are shown in Table \ref{tab15}. The specific parameters considered for the simulation of the IMPLY-based blocks are listed in Table \ref{tab15}, too. The values reported in Table \ref{tab15} are contemplated based on the parameters reported in \cite{ref2, ref4, ref6, ref39, ref40, ref41}. LTspice has been used to simulate the IMPLY-based logical and arithmetic blocks described in this paper.

\begin{table}[t]
	\centering
	\caption{Setup values of IMPLY logic and VTEAM model \cite{ref2, ref4, ref6}.}
	\scalebox{0.825}{
		\begin{tabular}{|c|c|c|c|c|c|}
			\hline
			Parameter & Value & Parameter & Value & Parameter & Value \\ \hline
			$v_{off}$ & 0.7 V & $v_{on}$ & -10 mV & $\alpha_{off}$ & 3 \\ \hline
			$\alpha_{on}$ & 3 & $R_{off}$ & 1 M$\Omega$ & $R_{on}$ & 10 k$\Omega$ \\ \hline
			$k_{on}$ & -0.5 $\frac{nm}{s}$ & $k_{off}$ & 1 $\frac{cm}{s}$ & $w_{off}$ & 0 nm \\ \hline
			$w_{on}$ & 3 nm & $w_{C}$ & 107 $pm$ & $a_{off}$ & 3 $nm$ \\ \hline
			$a_{on}$ & 0 $nm$ & $v_{set}$ & 1 V & $v_{reset}$ & 1 V \\ \hline
			$v_{cond}$ & 900 mV & $R_{G}$ & 40 K$\Omega$ & $t_{pulse}$ & 30 $\mu$s \\ \hline
	\end{tabular}}
	\label{tab15}
\end{table}

To analyze the functionality of the simulated circuits, note that, since all logic and arithmetic blocks are stateful, the input and output logical values are determined by the memristors' resistance. The maximum and minimum resistance values of the memristor are equal to logic '0’ and logic '1’, respectively. It should be noted that the functionality of IMPLY-based implementation algorithms must be evaluated by considering all possible input states. First, the input values must be initialized to the input memristors. Then the implementation algorithm runs for each state, and the correctness of the output is evaluated across all states against the intended resistance value of the output memristor. The number of computational steps, which is proportional to the computational delay, determines the performance. The energy consumption is computed using the method described in \cite{ref2, ref4, ref6, ref39, ref40, ref41, ref42}. For each input state, the energy consumption of each memristor in the row or column of the array is calculated, and the average over all states is reported as the estimated energy consumption. The number of memristors required is another criterion examined. In the following, the energy consumption, number of computational steps, and number of memristors required for each basic logic and arithmetic unit are calculated after verification. Then, the circuit evaluation criteria for the IMPLY-based Trivium and Grain-128a stream ciphers are reported in two scenarios. In the first scenario for each of the two stream ciphers, conventional algorithms are used without the improvements detailed in the previous section, whereas in the second scenario, the proposed methods are used to improve the circuit evaluation criteria.

\begin{figure}[h]
	\centering
	\includegraphics[scale=0.19]{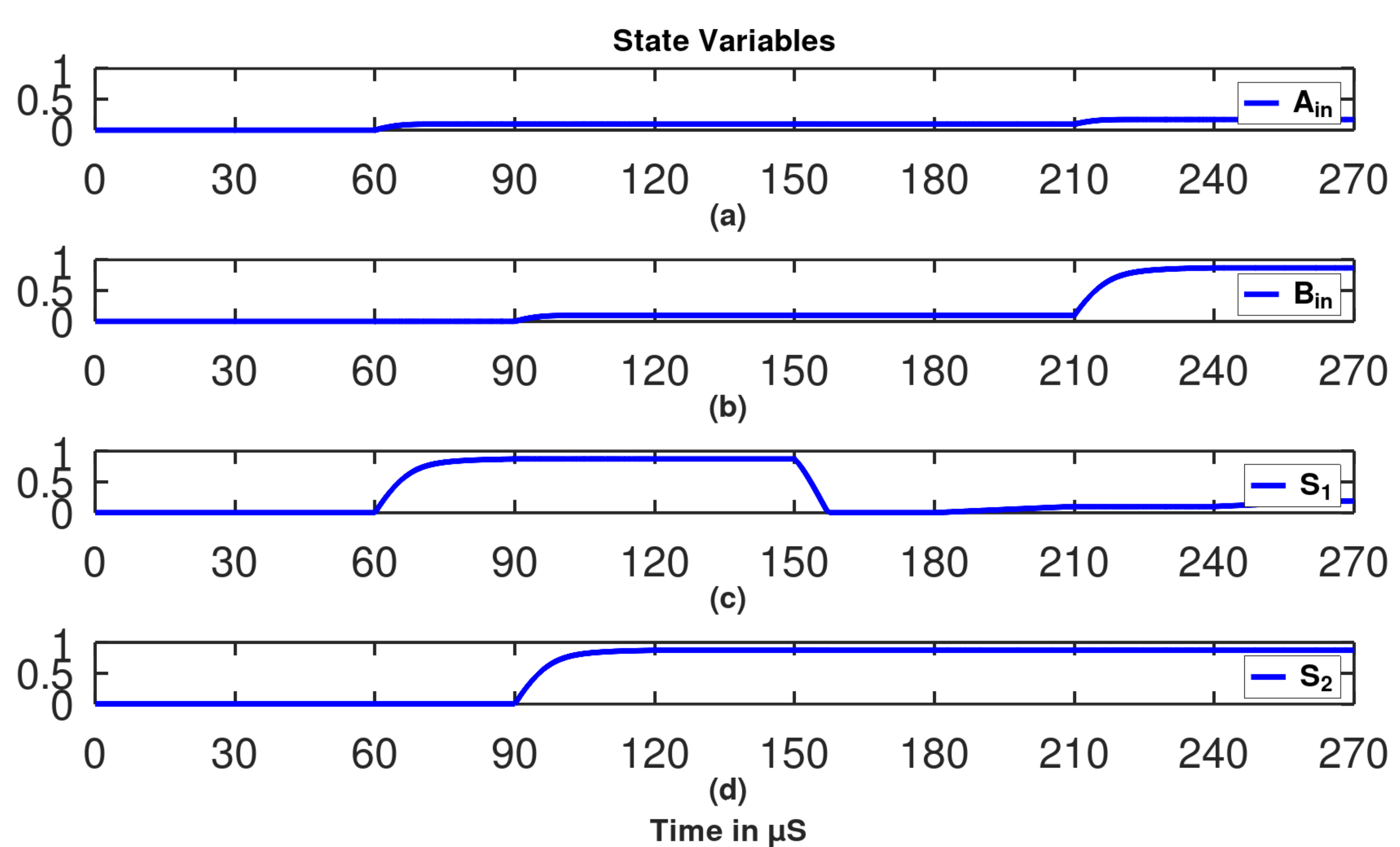}
	\caption{The output waveform of a destructive IMPLY-based serial two-input XOR gate introduced in \cite{ref6} for input state ``00".}
	\label{fig5}
\end{figure}

\begin{figure}[t!]
	\centering
	\includegraphics[scale=0.15]{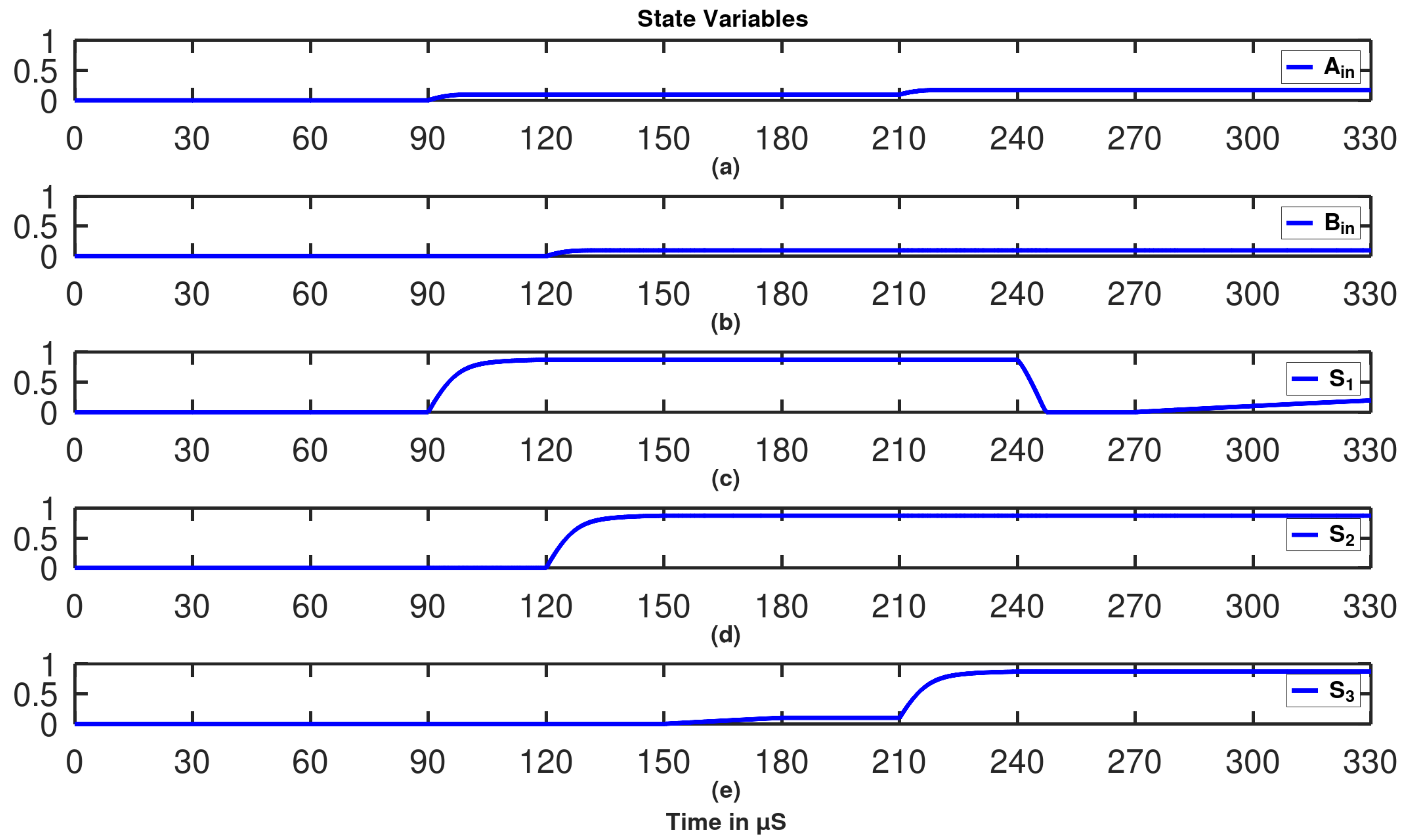}
	\caption{The output waveform of a non-destructive IMPLY-based serial two-input XOR gate for input state ``00".}
	\label{fig6}
\end{figure}

\begin{figure}[!]
	\centering
	\includegraphics[scale=0.15]{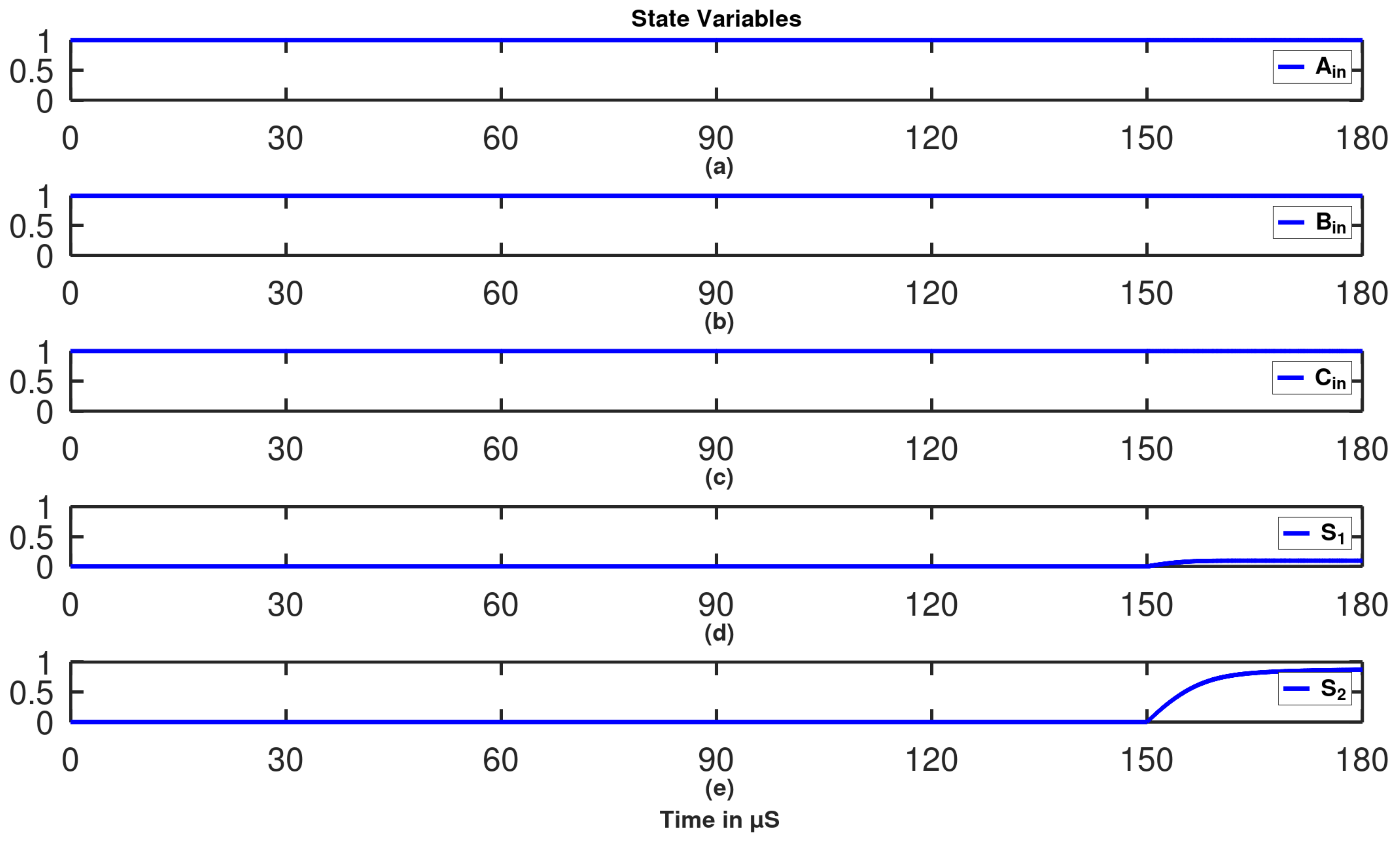}
	\caption{The output waveform of a IMPLY-based serial three-input AND gate for input state ``111".}
	\label{fig7}
\end{figure}

As mentioned, the functionality of the implementation algorithms for all IMPLY-based logic gates and basic arithmetic circuits used in the design of the Trivium and Grain-128a ciphers has been examined by considering all possible input states and applying the model in the LTSPICE simulator. Two-input XOR gates (according to the algorithms of Tables \ref{tab3} and \ref{tab4}), three-input XOR gate with the ability to retain input values, two-input, three-input, and four-input AND gates, buffers, and inverters are the circuits that are evaluated. The output waveforms of three gates, as examples of simulation results, are shown in Figures \ref{fig5}-\ref{fig7}. In Figure \ref{fig5}, the output waveform of a two-input XOR logic gate in which the logic value of one of the input memristors is changed is plotted. The inputs are assumed to be equal to ``00”, and the output is stored in the output memristor (the first work memristor, $s_{1}$) after nine computational cycles. According to Table \ref{tab15} and the reported value of $t_{step}$, the output is computed and stored in 240-270 $\mu$s. The analysis of the functionality of the other two-input XOR gate mentioned in Table \ref{tab3}, which computes the output in the last step of the implementation algorithm (300-330 $\mu$s), is shown in Figure \ref{fig6}. The inputs of this gate equal ``00”, and the input values stored in the memristors remain unchanged after the execution of the implementation algorithm. The output is stored in the first work memristor ($s_{1}$) in the final step. In Figure \ref{fig7}, the simulation result of the three-input AND gate ($A_{in}B_{in}C_{in}$=``111") is calculated in six computational steps according to (\ref{eq11}) in the time interval 0 to 180 $\mu$s. The output is obtained in the last computational step (150-180 $\mu$s) and stored in the second work memristor ($s_{2}$).

After introducing the applied model, familiarizing with the simulation method and circuit evaluation criteria, and presenting a number of simulation outputs, the results of the circuit evaluation criteria for the IMPLY-based basic logic gates required in the implementation of the Trivium and Grain-128a, lightweight stream ciphers, are summarized in Table \ref{tab16}. The numbers reported in Table \ref{tab16} are widely used in estimating the circuit evaluation criteria.

\begin{table}[t]
\centering
\caption{Summary of the circuit evaluation criteria for the IMPLY-based basic logic gates required in the implementation of the Trivium and Grain-128a.}
\scalebox{0.825}{
\begin{tabular}{|c|c|c|c|}
\hline
\textbf{Gate} & \textbf{No. of} & \textbf{No. of computational} & \textbf{Estimated energy} \\
&\textbf{Memristors} & \textbf{steps} & \textbf{consumption ($n$J)} \\ \hline
Inverter & 2 & 2 & 0.1291 \\ \hline
Buffer & 3 & 4 & 0.269 \\ \hline
2-input AND & 4 & 5 & 0.3833 \\ \hline
3-input AND & 5 & 6 & 0.5025 \\ \hline
4-input AND & 6 & 11 & 0.9131 \\ \hline
2-input XOR & 4 & 9 & 0.7426 \\
(destructive) \cite{ref6} & & & \\ \hline
2-input XOR & 5 & 11 & 0.9146 \\
(non-destructive) & & & \\ \hline
3-input XOR & 6 & 20 & 1.711 \\ \hline
\end{tabular}}
\label{tab16}
\end{table}

Subsection \ref{sec331} explains the implementation details of the Trivium cipher using the IMPLY method. The previous paragraphs of this subsection also mentioned the details of the circuit-level implementation of the basic building blocks of this cipher. Tables \ref{tab17} and \ref{tab18} specify the number of required computational units, the number of computational steps, and the estimated energy consumption for the initialization and keystream generation stages of the Trivium cipher, respectively.

The initialization phase of the Trivium cipher requires 293 memristors, including 288 input/output memristors and five work memristors, and the number of memristors is not increased in all cycles. Overall, the initialization phase of the IMPLY-based Trivium cipher takes 797266 computational steps and consumes 53.4731 $\mu$J. Each round of generating a keystream bit also requires the same number of memristors, takes 710 computational steps, and consumes 47.8983 $n$J. So, if the number of keystream bits required by the user is considered to be `$n$', the total number of computational steps and energy consumption estimation in the two phases are ``$710 \times n+797266$” steps and ``$0.0478 \times n+53.4731$” $\mu$J, respectively.

\begin{table}[h]
\centering
\caption{Details of the number of computational blocks, computational steps, and estimated energy consumption of the initialization stage of the serial IMPLY-based Trivium cipher.}
\scalebox{0.825}{
\begin{tabular}{|c|c|c|c|}
\hline
\textbf{Block} & \textbf{No. of} & \textbf{No. of} & \textbf{Estimated energy} \\
&\textbf{blocks} & \textbf{computational steps} & \textbf{consumption ($\mu$J)} \\ \hline
2-input XOR \cite{ref6} & 3456 & 17280 & 1.3246 \\ \hline
2-input AND & 10386 & 93312 & 7.699 \\ \hline
Inverter in the proposed & 103637 & 207274 & 13.3795 \\
algorithm (shift register $A$) & & & \\ \hline
Buffer in the proposed & 3499 & 13996 & 0.9412 \\
algorithm (shift register $A$) & & & \\ \hline
Inverter in the proposed & 92196 & 184392 & 11.9025 \\
algorithm (shift register $B$) & & & \\ \hline
Buffer in the proposed & 4572 & 18288 & 1.2298 \\
algorithm (shift register $B$) & & & \\ \hline
Inverter in the proposed & 124382 & 248764 & 16.0577 \\
algorithm (shift register $C$) & & & \\ \hline
Buffer in the proposed & 3490 & 13960 & 0.9388 \\
algorithm (shift register $C$) & & & \\ \hline
Buffer in the conventional & 107136 & 428544 & 28.8195 \\
algorithm (shift register $A$) & & & \\ \hline
Buffer in the conventional & 96768 & 387072 & 26.0305 \\
algorithm (shift register $B$) & & & \\ \hline
Buffer in the conventional & 127872 & 511488 & 34.3975 \\
algorithm (shift register $C$) & & & \\ \hline
\end{tabular}}
\label{tab17}
\end{table}

\begin{table}[h]
\centering
\caption{Details of the number of computational blocks, computational steps, and estimated energy consumption of the keystream generation stage of the serial IMPLY-based Trivium cipher.}
\scalebox{0.825}{
\begin{tabular}{|c|c|c|c|}
\hline
\textbf{Block} & \textbf{No. of} & \textbf{No. of} & \textbf{Estimated energy} \\
&\textbf{blocks} & \textbf{computational steps} & \textbf{consumption ($n$J)} \\ \hline
2-input XOR \cite{ref6} & 11 & 99 & 8.1686 \\ \hline
2-input AND & 3 & 15 & 1.1499 \\ \hline
Inverter in the  & 278 & 556 & 35.8898 \\
proposed algorithm  & & & \\ \hline
Buffer in the  & 10 & 40 & 2.69 \\
proposed algorithm & & & \\ \hline
Buffer in the  & 288 & 1152 & 77.472 \\
conventional algorithm & & & \\ \hline
\end{tabular}}
\label{tab18}
\end{table}

The circuit evaluation criteria for the Grain-128a cipher can be examined in the same way as for the Trivium cipher. The ``pre-initialization” phase of this cipher lasts 256 cycles, and the keystream is generated from the $257^{th}$ cycle. The number of memristors and logical units, the computational steps, and the estimated energy consumption of the units applied in the ``pre-initialization” and keystream generation phases are reported in Tables \ref{tab19} and \ref{tab20}, respectively.

\begin{table}[h]
\centering
\caption{Details of the number of computational blocks, computational steps, and estimated energy consumption of the ``pre-initialization" stage of the serial IMPLY-based Grain-128a cipher.}
\scalebox{0.825}{
\begin{tabular}{|c|c|c|c|}
\hline
\textbf{Block} & \textbf{No. of} & \textbf{No. of} & \textbf{Estimated energy} \\
&\textbf{blocks} & \textbf{computational steps} & \textbf{consumption ($\mu$J)} \\ \hline
2-input XOR & 7936 & 71424 & 5.8932 \\
(destructive) \cite{ref6} & & & \\ \hline
2-input XOR & 768 & 8448 & 0.7024 \\
(non-destructive) & & & \\ \hline
2-input AND & 2816 & 14080 & 1.0793 \\ \hline
3-input AND & 768 & 4608 & 0.3859 \\ \hline
4-input AND & 256 & 2816 & 0.2237 \\ \hline
Inverter in the & 31195 & 62390 & 4.0272 \\
proposed algorithm (LFSR) & & & \\ \hline
Buffer in the & 1573 & 6292 & 0.4231 \\
proposed algorithm (LFSR) & & & \\ \hline
Buffer in the & 32768 & 131072 & 8.8145 \\
conventional algorithm (LFSR) & & & \\ \hline
Inverter in the & 27650 & 55300 & 3.5696 \\
proposed algorithm (NFSR) & & & \\ \hline
Buffer in the & 5118 & 20472 & 1.3767 \\
proposed algorithm (NFSR) & & & \\ \hline
Buffer in the & 32768 & 131072 & 8.8145 \\
conventional algorithm (NFSR) & & & \\ \hline
\end{tabular}}
\label{tab19}
\end{table}

\begin{table}[!]
\centering
\caption{Details of the number of computational blocks, computational steps, and estimated energy consumption of the keystream generation stage of the serial IMPLY-based Grain-128a cipher.}
\scalebox{0.825}{
\begin{tabular}{|c|c|c|c|}
\hline
\textbf{Block} & \textbf{No. of} & \textbf{No. of} & \textbf{Estimated energy} \\
&\textbf{blocks} & \textbf{computational steps} & \textbf{consumption ($n$J)} \\ \hline
2-input XOR & 29 & 261 & 21.5354 \\
(destructive) \cite{ref6} & & & \\ \hline
2-input XOR & 3 & 33 & 2.7438 \\
(non-destructive) & & & \\ \hline
2-input AND & 11 & 55 & 4.2163 \\ \hline
3-input AND & 3 & 18 & 1.5075 \\ \hline
4-input AND & 1 & 11 & 0.9131 \\ \hline
Inverter in the & 122 & 244 & 15.7502 \\
proposed algorithm (LFSR) & & & \\ \hline
Buffer in the & 6 & 24 & 1.614 \\
proposed algorithm (LFSR) & & & \\ \hline
Buffer in the & 128 & 512 & 34.432 \\
conventional algorithm (LFSR) & & & \\ \hline
Inverter in the & 108 & 216 & 13.9428 \\
proposed algorithm (NFSR) & & & \\ \hline
Buffer in the & 20 & 80 & 5.38 \\
proposed algorithm (NFSR) & & & \\ \hline
Buffer in the & 128 & 512 & 34.432 \\
conventional algorithm (NFSR) & & & \\ \hline
\end{tabular}}
\label{tab20}
\end{table}

Two hundred and sixty two memristors participate in the ``pre-initialization” phase of the Grain-128a cipher. This phase requires 245830 computational steps and estimates an energy consumption of 17.6811 $\mu$J. Each round of keystream generation requires 942 computational steps and 263 memristors. The energy consumption estimation of the second stage of the Grain-128a is 0.0666 $\mu$J. The computed values can be generalized for an n-bit keystream. To produce an n-bit keystream of this cipher, ``$942\times n+245830$” computational steps and 263 memristors are required. The energy consumption estimate for this cipher in generating an n-bit keystream is also ``$0.0666\times n+17.6811$” $\mu$J.

A comprehensive comparison of the circuit evaluation criteria for the Trivium and Grain-128a ciphers in the mentioned scenarios is presented in Table \ref{tab21}. The number of memristors required by both ciphers to generate an n-bit keystream is 293 and 263, respectively. The number of computational steps and energy consumption of the Trivium cipher applying the proposed shift register’s implementation algorithm, considering the generation of 10,000 bits, are 18\% and 22\% less than that of the Grain-128a cipher, respectively, while by increasing the number of output bits by a factor of 10, the number of computational steps and energy consumption estimation are improved by 24\% and 27\%.

According to Table \ref{tab21}, it can be concluded that the computational delay and energy consumption of the Trivium cipher are lower than those of Grain-128a, and it is preferred over Grain-128a if computational delay and energy consumption are of higher priority for the designer. Grain-128a has higher priority than Trivium if the designers' main goal is to use a cipher with a 128-bit key and the circuit-level evaluation metrics are second-tier. As the number of bits tends to infinity, the number of computational steps and the energy consumption of the improved IMPLY-based Trivium and Grain-128a lightweight stream ciphers based on the proposed shift register algorithm, are improved by 38\%, 42\%, 45\%, and 32\% compared to the conventional design, respectively.

\begin{table}[h]
\centering
\caption{Comprehensive comparison of circuit evaluation metrics of IMPLY-based serial Trivium and Grain-128a ciphers.}
\scalebox{0.8}{
\begin{tabular}{|c|c|c|c|c|c|c|}
\hline
\textbf{Lightweight} & \textbf{No. of} & $n=10000$ & $n=100000$& \textbf{Estimated energy} & $n=10000$ & $n=100000$ \\
\textbf{stream cipher} & \textbf{computational steps} & & & \textbf{consumption ($\mu$J)} & & \\ \hline
Trivium (conventional) & $1152 \times n+1437696$ & 12957696 & 116637696 & $0.0867 \times n+98.2711$ & 965.2711 & 8768.2711 \\ \hline
Trivium (proposed) & $710 \times n+797266$ & 7897266 & 71797266 & $0.0478 \times n+53.4731$ & 531.4731 & 4833.4731 \\ \hline
Grain-128a (conventional) & $1646 \times n+363520$ & 16823520 & 164963520 & $0.09878 \times n+25.9135$ & 1013.7135 & 9903.9135 \\ \hline
Grain-128a (proposed) & $942 \times n+245830$ & 9665830 & 94445830 & $0.0666 \times n+17.6811$ & 683.6811 & 6677.6811 \\ \hline
\end{tabular}}
\label{tab21}
\end{table}

\subsection{Application-level simulation} \label{sec42}
GNU Octave has been used to simulate the behavior of the serial IMPLY-based Trivium and Grain-128a stream ciphers in cryptography and steganography. All the logic gates and logical/arithmetic units have been simulated behaviorally using the algorithms mentioned in the previous sections, and their functionality has been validated. The proposed IMPLY-based shift register algorithm has also been considered in the behavioral simulation of both ciphers' implementations. The core structures of each cipher were implemented using basic logical circuits designed with the IMPLY method. After verifying the correctness of the cores' functionality, both ciphers were implemented in their two main stages of operation: initialization and keystream generation. In the final code, the desired keystream was generated, and encryption and decryption operations were performed based on the structures of both ciphers.

One goal of implementing lightweight ciphers based on the IMPLY design method is to evaluate their functionality in steganography applications. Hiding data bits in the LSB of each pixel in an image is one of the most common and low-cost methods of steganography \cite{ref43}. Therefore, to examine the functionality of the implemented circuits in this application, random messages were encrypted using the designed IMPLY-based Trivium and Grain-128a, and the output bits were inserted one bit at a time into the LSBs of the carrier images. The output messages from the previous step were also recovered from the images, and after comparing the recovered data with the original data, it can be deduced that the entire process was performed correctly.

Assessing image quality metrics such as Peak Signal to Noise Ratio (PSNR) and comparing the histograms of the steganographic output with the original image, ensures the reliability of the process. The human eye cannot distinguish the original image from the steganographic output if the PSNR is above 30 dB \cite{ref44}. Hence, if the PSNR is higher than 30 dB, it can be concluded that the steganographic output is acceptable, as it is not possible to easily distinguish the two images from each other, and even human recognition is impossible. By drawing histograms of the original image and the output of the steganography process, and analyzing the outputs, the user can determine how well this technique works for hiding data.

\begin{figure}[h]
\centering
\subcaptionbox{}{ \includegraphics[width=0.2\textwidth]{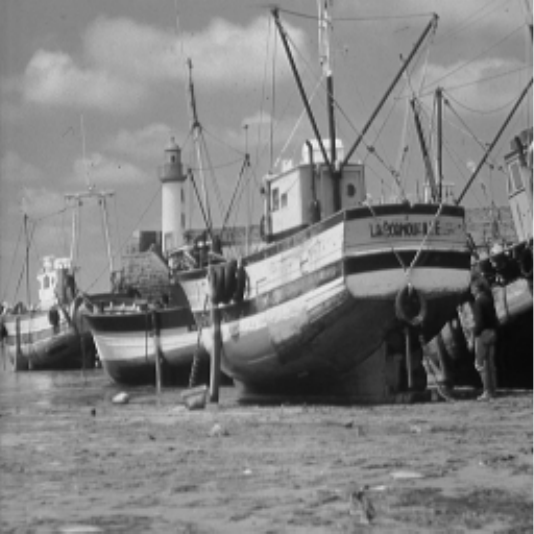} }
\subcaptionbox{}{ \includegraphics[width=0.2\textwidth]{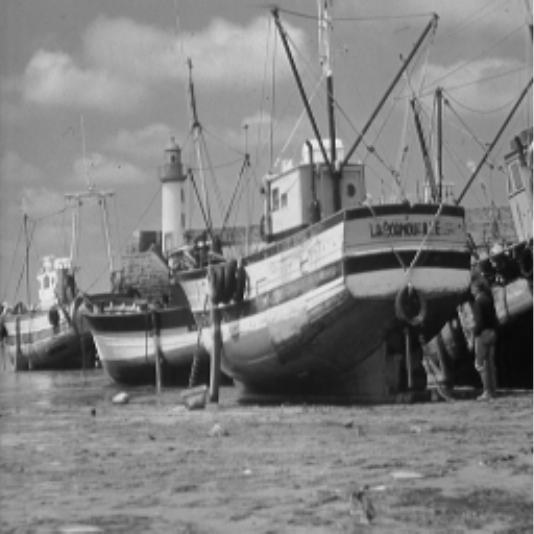} }
\subcaptionbox{}{ \includegraphics[width=0.2\textwidth]{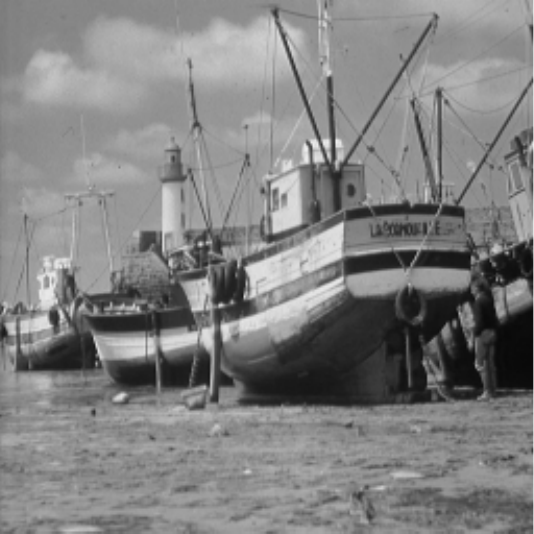} }
\caption{Functionality analysis of IMPLY-based lightweight stream ciphers in steganography:(a) Cover image (``boat"), (b) Trivium-based stego image, and (c) Grain-128a stego image.}
\label{fig9}
\end{figure}

\begin{figure}[h]
\centering
\subcaptionbox{}{ \includegraphics[width=0.2\textwidth]{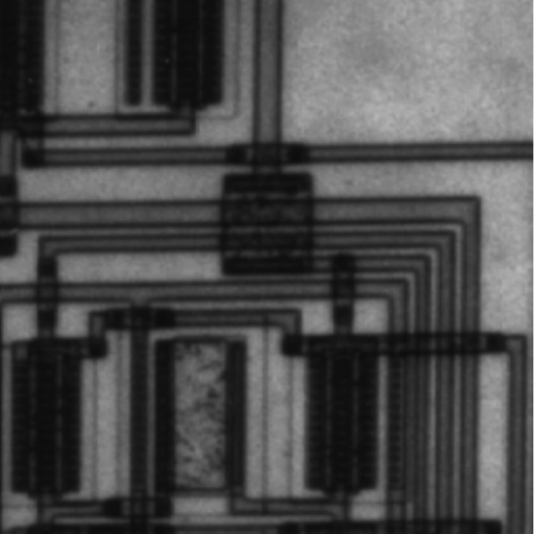} }
\subcaptionbox{}{ \includegraphics[width=0.2\textwidth]{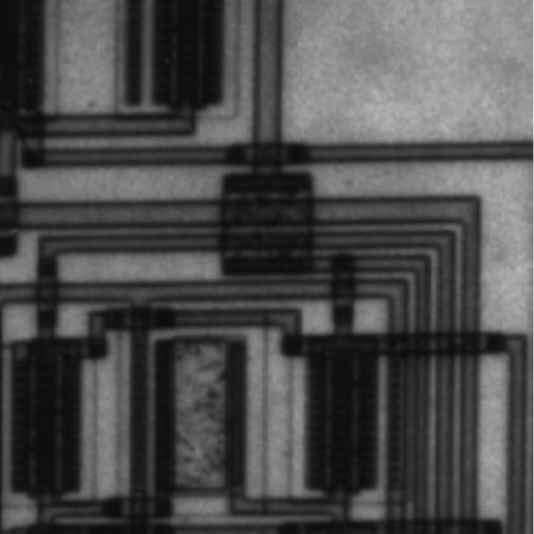} }
\subcaptionbox{}{ \includegraphics[width=0.2\textwidth]{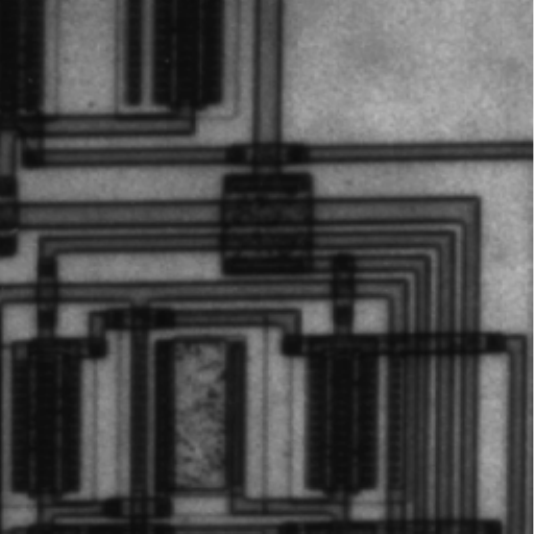} }
\caption{Functionality analysis of IMPLY-based lightweight stream ciphers in steganography:(a) Cover image (``circuit"), (b) Trivium-based stego image, and (c) Grain-128a stego image.}
\label{fig10}
\end{figure}

According to the method applied to the behavioral simulation discussed in the previous paragraphs, five random messages of varying lengths were generated and, after processing, placed in the LSBs of the pixels of five standard 256$\times$256 grayscale images. The output of the steganography process consists of two images, along with the standard grayscale image. The first and second output images are generated from the outputs of the lightweight stream ciphers Trivium and Grain-128a, respectively. Two sets of outputs from the steganography application are shown in Figures \ref{fig9} and \ref{fig10}. The PSNR values of this application is tabulated in Table \ref{tab22}.

\begin{table}[!]
\centering
\caption{Evaluation of the PSNR values of the stego images generated using the IMPLY-based Trivium and Grain-128a ciphers.}
\scalebox{0.825}{
\begin{tabular}{|c|c|c|}
\hline
\textbf{Cover} & \textbf{Trivium-based stego image} & \textbf{Grain-128a-based stego image} \\
\textbf{image} & \textbf{PSNR value (dB)} & \textbf{PSNR value (dB)} \\ \hline
Boat & 71.642 & 71.966 \\ \hline
Cameraman & 69.822 & 69.668 \\ \hline
Circuit & 69.715 & 69.783 \\ \hline
Rice & 73.199 & 73.178 \\ \hline
Walkbridge & 65.139 & 65.258 \\ \hline
\end{tabular}}
\label{tab22}
\end{table}

\begin{figure}[t!]
\centering
\begin{subfigure}{0.7\textwidth}
\centering
\includegraphics[width=0.7\linewidth]{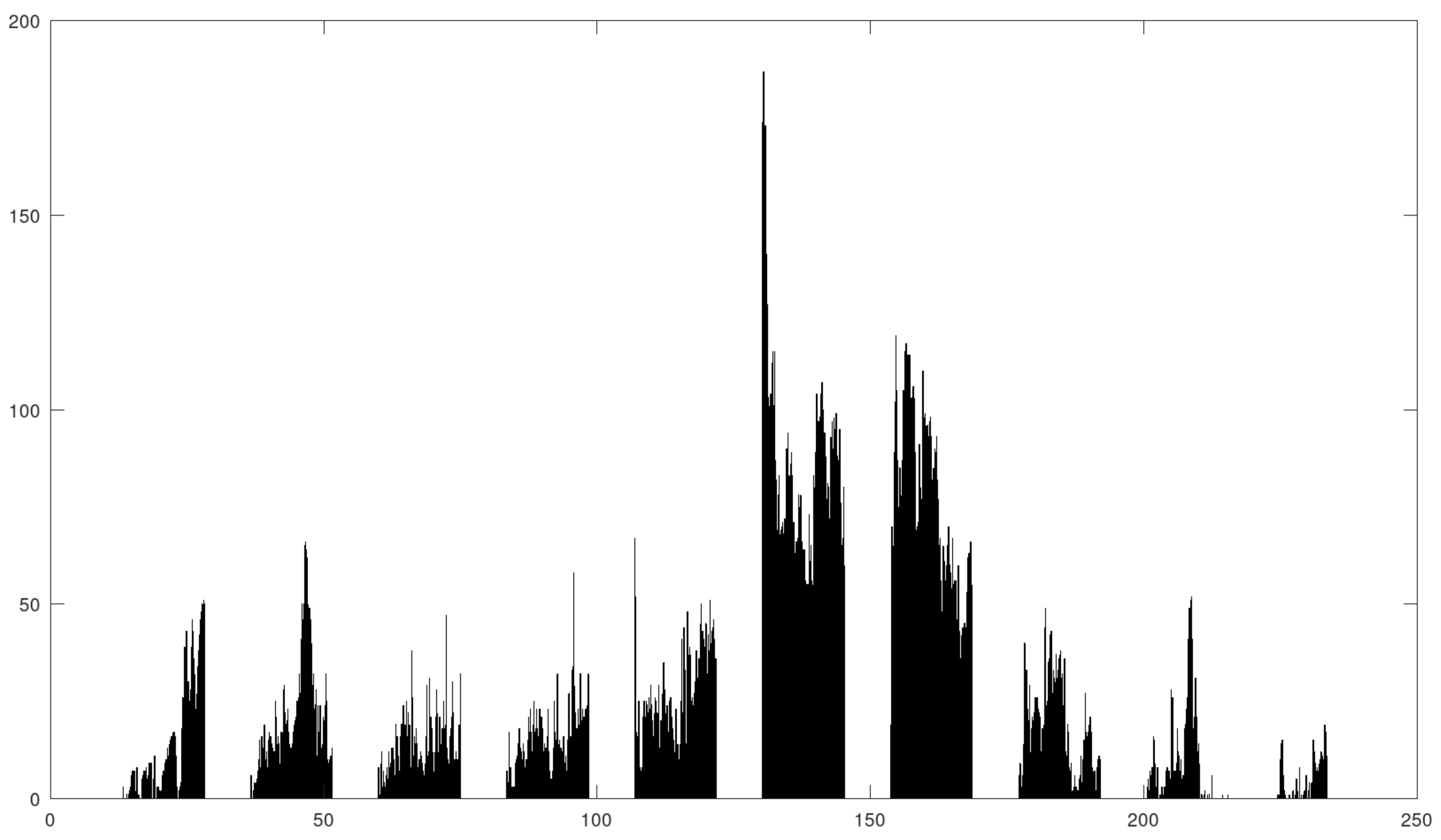}
\caption{}
\end{subfigure}
\begin{subfigure}{0.35\textwidth}
\centering
\includegraphics[width=0.7\linewidth]{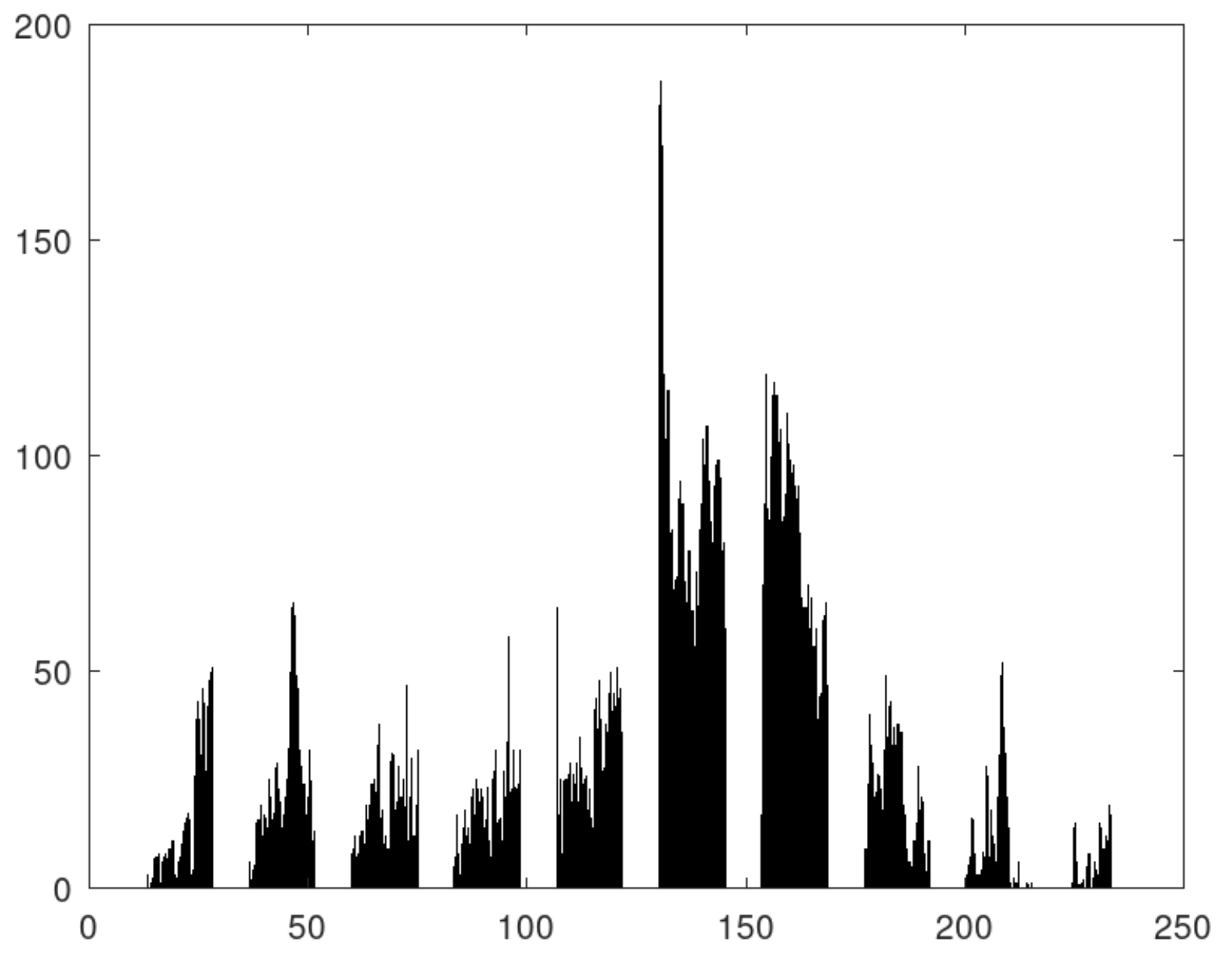}
\caption{}
\end{subfigure}
\hspace{0pt}
\begin{subfigure}{0.35\textwidth}
\centering
\includegraphics[width=0.7\linewidth]{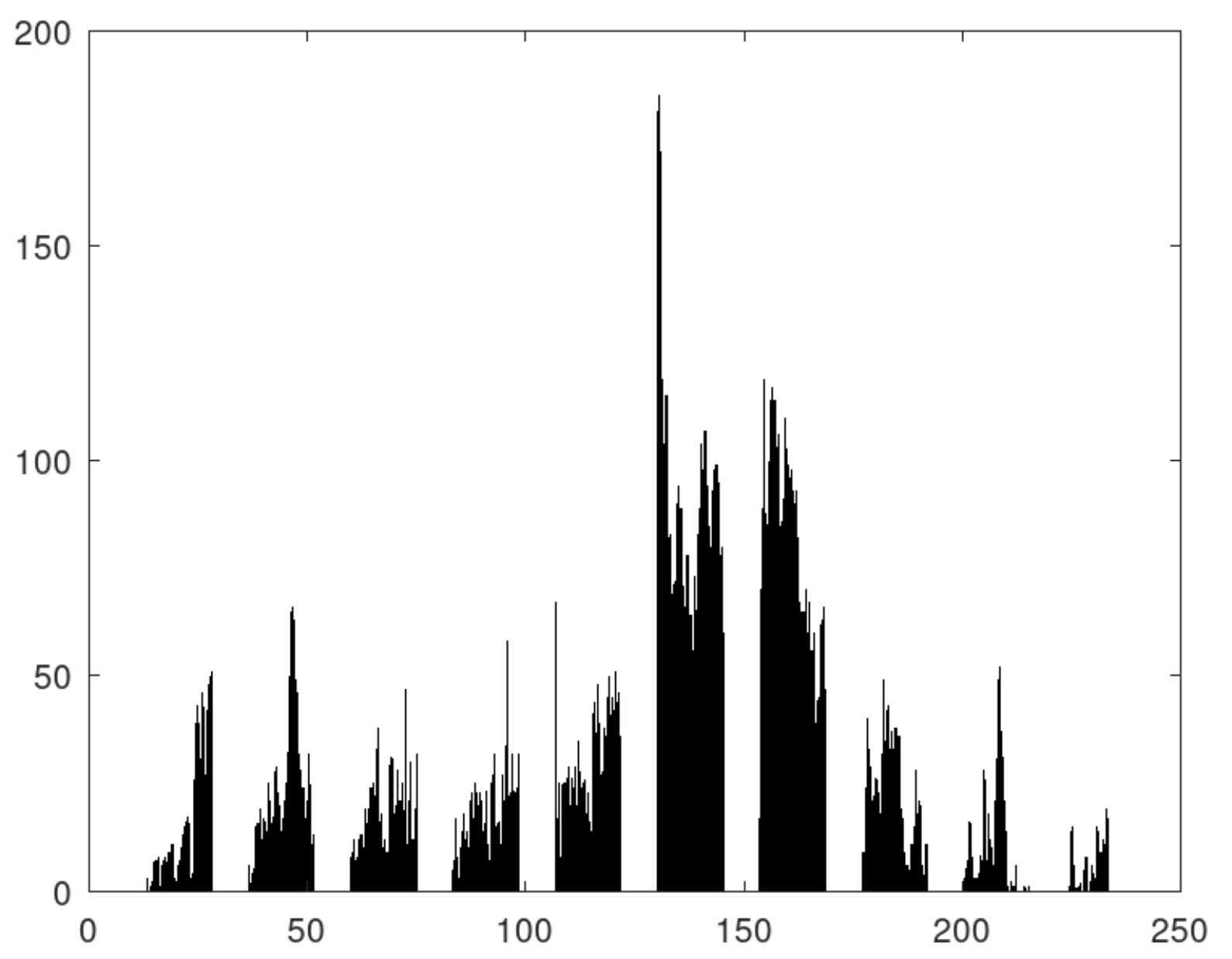}
\caption{}
\end{subfigure}
\caption{Histograms of (a) Figure \ref{fig9}(a), (b) Figure \ref{fig9}(b), and (c) Figure \ref{fig9}(c).}
\label{fig11}
\end{figure}

\begin{figure}[t!]
\centering
\begin{subfigure}{0.7\textwidth}
\centering
\includegraphics[width=0.7\linewidth]{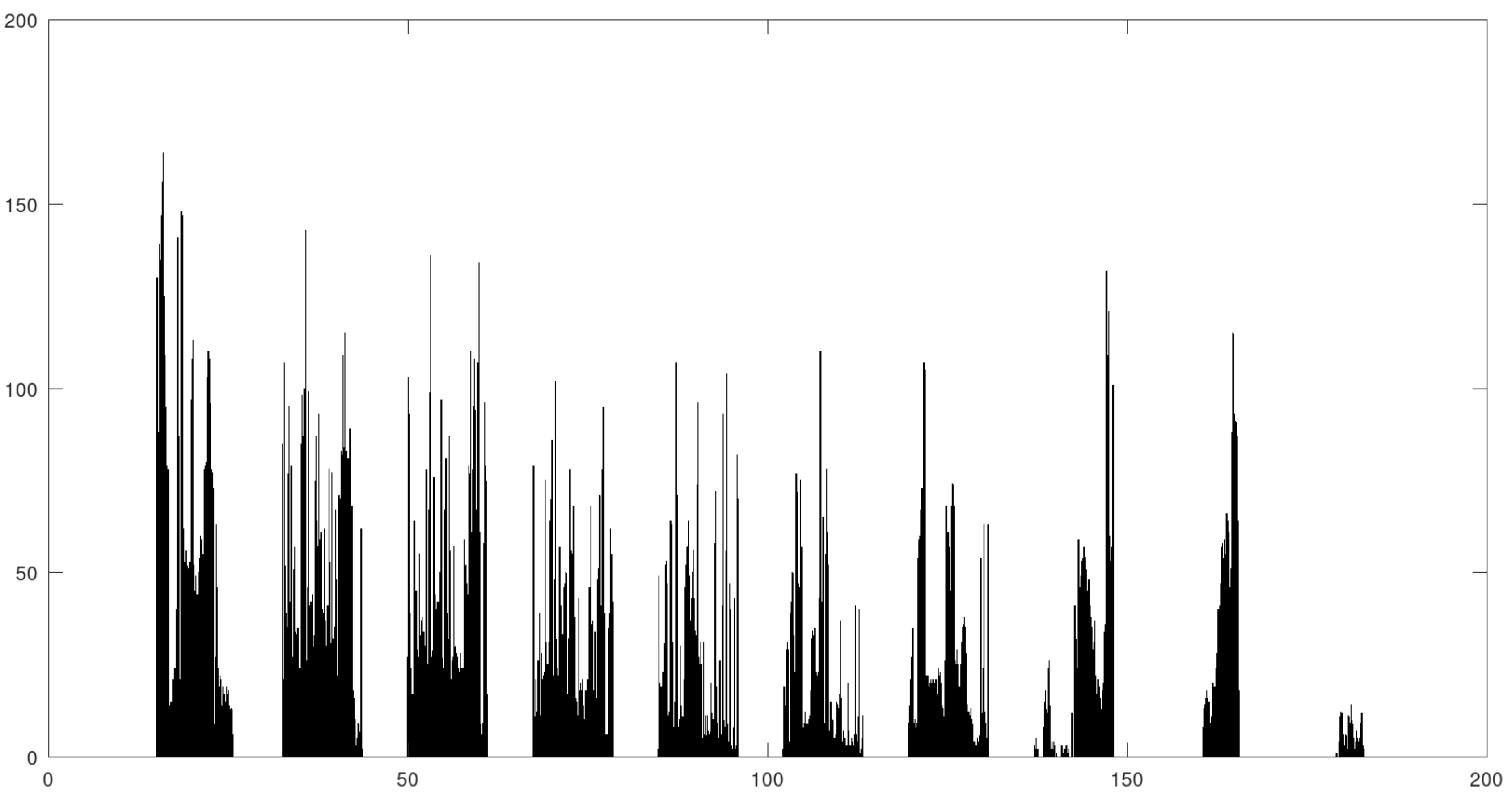}
\caption{}
\end{subfigure}
\begin{subfigure}{0.35\textwidth}
\centering
\includegraphics[width=0.7\linewidth]{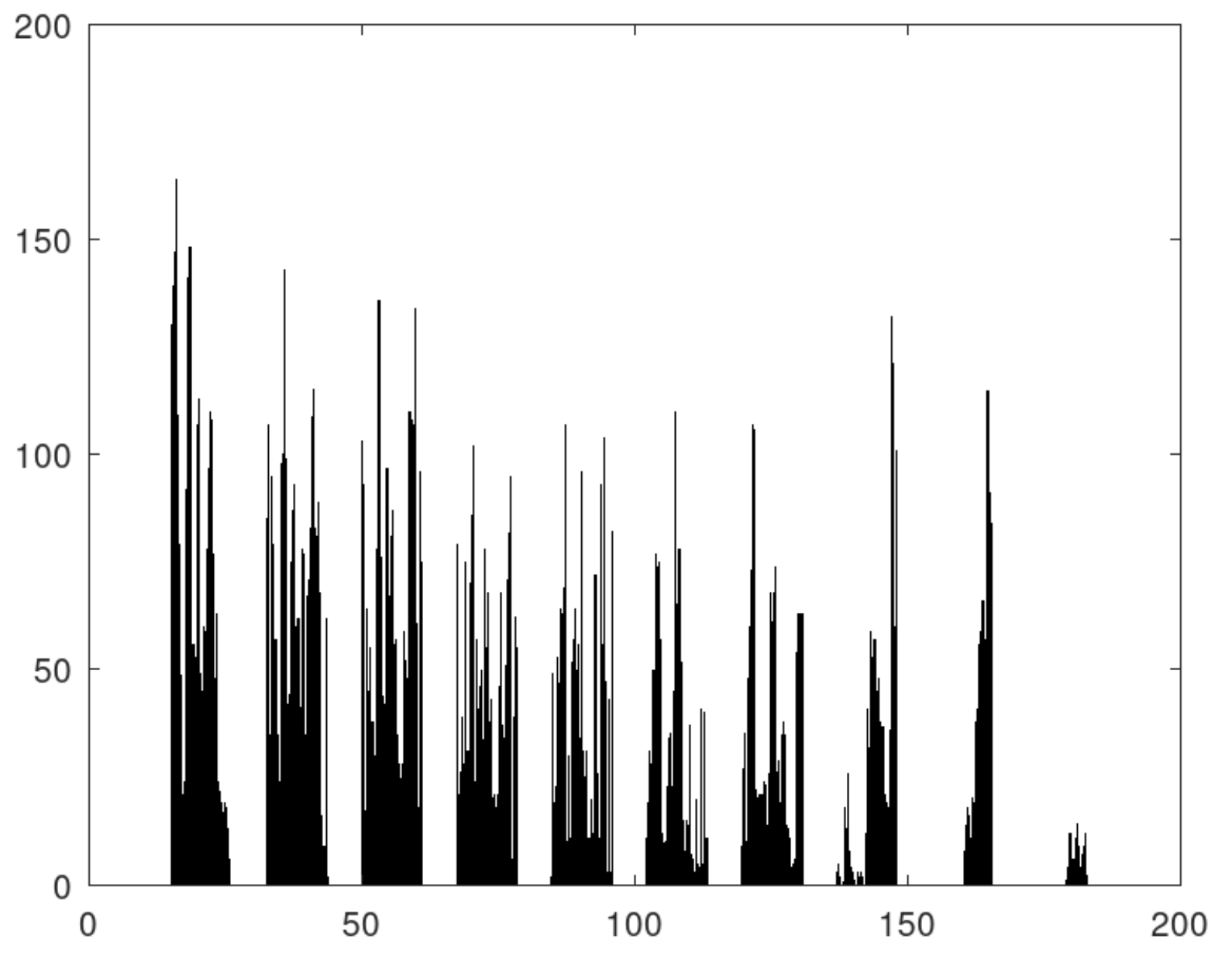}
\caption{}
\end{subfigure}
\hspace{0pt}
\begin{subfigure}{0.35\textwidth}
\centering
\includegraphics[width=0.7\linewidth]{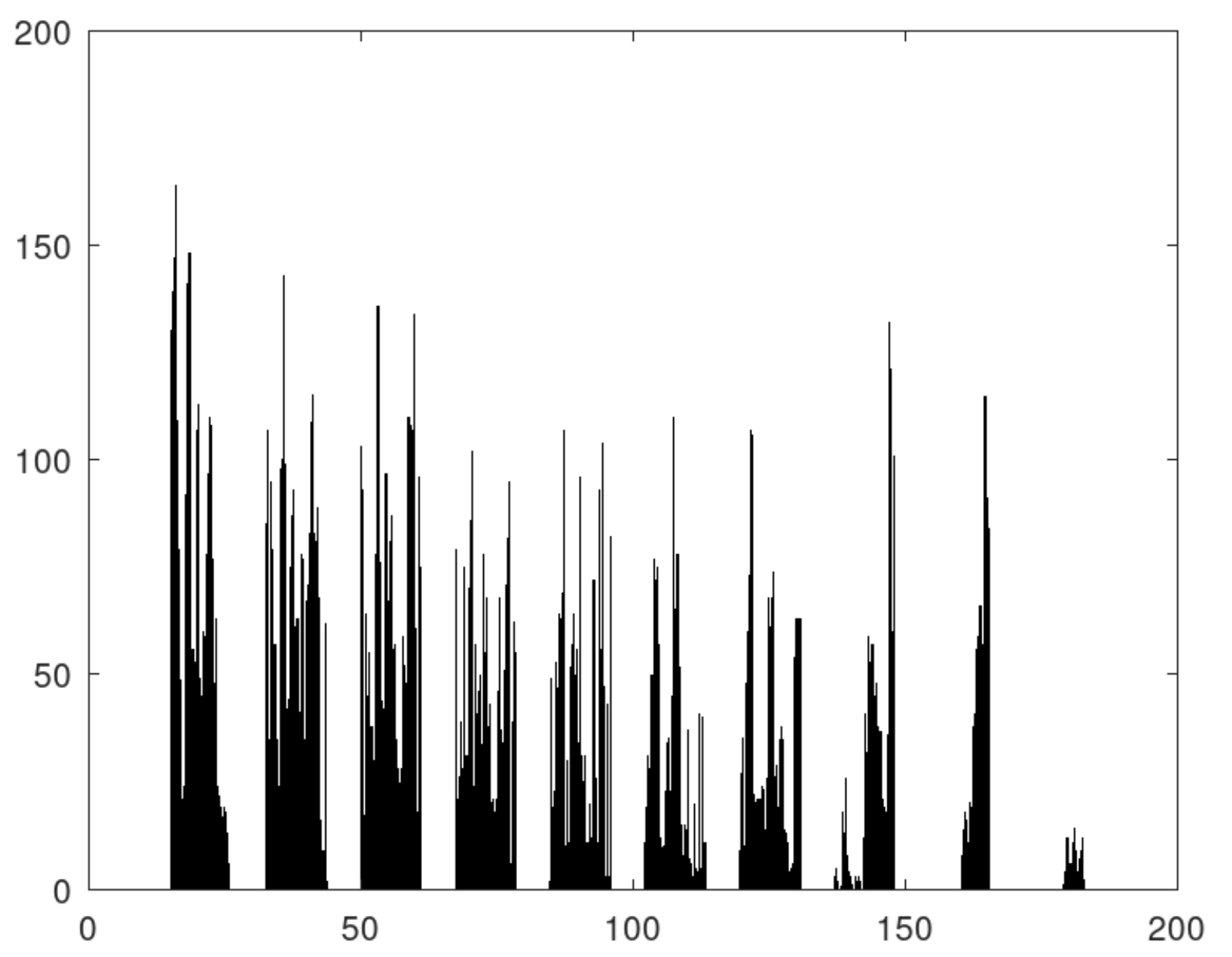}
\caption{}
\end{subfigure}
\caption{Histograms of (a) Figure \ref{fig10}(a), (b) Figure \ref{fig10}(b), and (c) Figure \ref{fig10}(c).}
\label{fig12}
\end{figure}

According to the results in Table \ref{tab22} and Figures \ref{fig9} and \ref{fig10}, it can be concluded that hiding the outputs of random messages of different lengths encrypted using the Trivium and Grain-128a ciphers in the LSBs of standard grayscale images is acceptable. As the number of encrypted bits increases, image quality decreases, but the results show a large gap between the output PSNR values and the acceptable image quality margin (30 dB). By increasing the number of bits, larger images can be used to improve image quality. Histograms corresponding to Figures \ref{fig9} and \ref{fig10} are also shown in Figures \ref{fig11} and \ref{fig12}; thus, the pixel distributions of the outputs of the steganography application can be compared and evaluated with the reference images.

\section{Conclusion} \label{sec5}
In this paper, two lightweight stream ciphers, Trivium and Grain-128a, are implemented using the serial IMPLY-based design method with a minimum number of memristors harmonized with the structure of the memristive crossbar array. In the redesigned ciphers, an attempt has been made to minimize the number of computational steps. Accordingly, in addition to investigating the possibility of overlapping steps in the implementation of basic logical and arithmetic blocks, by replacing the buffers required in the structure of shift registers of the architectures of the Trivium and Grain-128a ciphers with a combination of inverters and buffers, this study attempts to significantly improve the circuit evaluation criteria, such as the number of computational steps and energy consumption. Applying the proposed method to redesign the shift register structures in these two lightweight stream ciphers has resulted in improvements of 38\%-42\% in the number of computational steps and 32\%-44\% in energy consumption compared to the conventional design. The improved structure of the IMPLY-based Trivium cipher generates the keystream faster and consumes less energy than Grain-128a, such that by generating 10,000 valid bits employing the improved IMPLY-based Trivium cipher, the number of computational steps and energy consumption are 18\% and 22\% less than circuit evaluation criteria of the IMPLY-based Grain-128a architecture, respectively. By examining the PSNR criterion and the histogram diagrams of the different output images from the steganography application, it can be concluded that the functionality of the designed circuits in this application was acceptable.

\subsection*{Acknowledgment}
During the preparation of this work, the authors utilized the ``Gemini 3.1 Pro" to improve the readability, grammar, and formal academic tone of the manuscript. After using this tool, the authors meticulously reviewed and edited the content. The authors take full responsibility for the scientific accuracy, originality, and overall integrity of the final publication.

\subsection*{Author Contributions}
\noindent \textbf{Seyed Erfan Fatemieh}: Conceptualization, Data Curation, Formal Analysis, Investigation, Methodology, Software, Validation, Visualization, Writing-original draft, Writing-review \& editing.

\noindent \textbf{Reza Shahdi Alizadeh}: Project Administration, Resources, Supervision, Validation, Writing-review \& editing.

\noindent \textbf{Esmail Zarezadeh}: Project Administration, Supervision, Validation, Writing-review \& editing.

\subsection*{Data Availability Statement}
Data is contained within the article.

\section*{Statements \& Declarations}
\subsection*{Competing interests}
The authors declare no conflict of interest.

\subsection*{Funding}
No funds, grants, or other support was received.

\bibliographystyle{ieeetr}
\bibliography{Manuscript}

\end{document}